\documentstyle[11pt]{article}

\if@twoside\oddsidemargin25mm
\else\oddsidemargin25mm\evensidemargin25mm\marginparwidth25mm\fi%
\newcommand{\be}{\begin{equation}}
\newcommand{\ee}{\end{equation}}
\newcommand{\bea}{\begin{eqnarray}}
\newcommand{\eea}{\end{eqnarray}}
\newcommand{\beasn}{\begin{sneqnarray}}
\newcommand{\eeasn}{\end{sneqnarray}}
\newcommand{\bref}[1]{(\ref{#1})}
\newcommand{\eps}{\epsilon}
\newcommand{\veps}{\varepsilon}

\newcommand{\ct}[1]{\cite{#1}}

\newcommand{\der}[2]{\frac{\partial #1}{\partial #2}}
\newcommand{\lder}[2]{\frac{\partial_l#1}{\partial #2}}
\newcommand{\rder}[2]{\frac{\partial_r#1}{\partial #2}}
\newcommand{\gh}[1]{{\cal #1}}

\newfont{\nice}{eufm10 scaled\magstep1}

\makeatletter

\def\restric#1#2{{\left. #1 \right|_{#2}}}
\def\dif{{\rm d}}
\def\deriv{\@ifnextchar[{\@deriv}{\@deriv[]}}
   \def\@deriv[#1]#2#3{\mathchoice%
{{\dif^{#1}#2\over\dif{#3}^{#1}}}{{\dif^{#1}#2/\dif{#3}^{#1}}}%
{{\dif^{#1}#2\over\dif{#3}^{#1}}}{{\dif^{#1}#2/\dif{#3}^{#1}}}}

\def\secteqno{\@addtoreset{equation}{section}%
\def\theequation{\thesection.\arabic{equation}}}

\def\endsecteqno{\def\theequation{\@ifundefined{chapter}%
{\arabic{equation}}{\thechapter.\arabic{equation}}}}

\newcounter{subequation}
\def\thesubequation{\alph{subequation}}
\def\sneqnarray{\stepcounter{equation}\let\@currentlabel=\theequation
\setcounter{subequation}{1}
\def\@eqnnum{{\rm (\theequation.\thesubequation)}}
\global\@eqcnt\z@\tabskip\@centering\let\\=\@eqncr\let\@@eqncr=\@@sneqncr
$$\halign to \displaywidth\bgroup\@eqnsel\hskip\@centering
 $\displaystyle\tabskip\z@{##}$&\global\@eqcnt\@ne
 \hskip 2\arraycolsep \hfil${##}$\hfil
 &\global\@eqcnt\tw@ \hskip 2\arraycolsep $\displaystyle\tabskip\z@{##}$\hfil
  \tabskip\@centering&\llap{##}\tabskip\z@\cr}
\def\endsneqnarray{\@@sneqncr\egroup $$\global\@ignoretrue}
\def\@@sneqncr{\let\@tempa\relax
   \ifcase\@eqcnt \def\@tempa{& & &}\or \def\@tempa{& &}
   \else \def\@tempa{&}\fi
     \@tempa \if@eqnsw\@eqnnum\stepcounter{subequation}\fi
     \global\@eqnswtrue\global\@eqcnt\z@\cr}

\def\nobiblabels{\def\@lbibitem[##1]##2{\@bibitem{##2}}}

\makeatother

\def\AP#1#2#3{ {{\sl Ann.\,Phys.\,}(N.Y.)\,}
    {\bf  {#1}} ({#2}) {#3}}
\def\CMP#1#2#3{ {\sl Commun.\,Math.\,Phys.\,}
    {\bf  {#1}} ({#2}) {#3}}
\def\IJMPA#1#2#3{ {\sl Int.\,J.\,Mod.\,Phys.\,}
    {\bf A{#1}} ({#2}) {#3}}

\def\NPB#1#2#3{ {\sl Nucl.\,Phys.\,}
    {\bf B{#1}} ({#2}) {#3}}

\def\PLB#1#2#3{ {\sl Phys.\,Lett.\,}
    {\bf B{#1}} ({#2}) {#3}}

\def\PRD#1#2#3{ {\sl Phys.\,Rev.\,}
    {\bf D{#1}} ({#2}) {#3}}

\newsavebox{\uuunit}
\sbox{\uuunit}
    {\setlength{\unitlength}{0.825em}
     \begin{picture}(0.6,0.7)
        \thinlines
        \put(0,0){\line(1,0){0.5}}
        \put(0.15,0){\line(0,1){0.7}}
        \put(0.35,0){\line(0,1){0.8}}
       \multiput(0.3,0.8)(-0.04,-0.02){12}{\rule{0.5pt}{0.5pt}}
     \end {picture}}
\newcommand {\unity}{\mathord{\!\usebox{\uuunit}}}

\secteqno

\begin{document}

\input FEYNMAN

\begin{titlepage}

\begin{flushright}

KUL-TF-95/3 \\
hep-th/9502140 \\
February 1995\\

\end{flushright}

\vspace{5mm}

\begin{center}

{\LARGE\bf Nonlocally Regularized Antibracket--Antifield Formalism
           and Anomalies in Chiral $W_3$ Gravity}

\vskip 2.cm

{\sc Jordi Par\'{\i}s$^\sharp$}

\vskip 0.5cm

\small{\it{Instituut voor Theoretische Fysica}}\\
\small{\it{Katholieke Universiteit Leuven}}\\
\small{\it{Celestijnenlaan 200D}}\\
\small{\it{B-3001 Leuven, Belgium}}\\
[2.5cm]

{\bf Abstract}

\end{center}

\begin{quote}

The nonlocal regularization method, recently proposed in
ref.\,\ct{emkw91,kw92,kw93}, is extended to general gauge theories
by reformulating it along the ideas of the antibracket-antifield
formalism. From the interplay of both frameworks a fully regularized
version of the field-antifield (FA) formalism arises, being able to deal
with higher order loop corrections and to describe higher order loop
contributions to the BRST anomaly. The quantum master equation, considered
in the FA framework as the quantity parametrizing BRST anomalies, is
argued to be incomplete at two and higher order loops and conjectured to
reproduce only the one-loop corrections to the $\hbar^p$ anomaly generated
by the addition of $O(\hbar^{k})$, $k<p$, counterterms.

Chiral $W_3$ gravity is used to exemplify the nonlocally regularized FA
formalism. First, the regularized one-loop quantum master equation is used
to compute the complete one-loop anomaly. Its two-loop order, however, is
shown to reproduce only the modification to the two-loop anomaly produced
by the addition of a suitable one-loop counterterm, thereby providing an
explicit verification of the previous statement for $p=2$.
The well-known universal two-loop anomaly, instead, is alternatively
obtained from the BRST variation of the nonlocally regulated effective
action. Incompleteness of the quantum master equation is thus concluded to
be a consequence of a naive derivation of the FA BRST Ward identity.

\vspace{10mm}

\hrule width 5.cm

{\small \noindent $^\sharp$ E-mail: Jordi.Paris@fys.kuleuven.ac.be}

\normalsize
\end{quote}

\end{titlepage}

\section{Introduction}

\hspace{\parindent}%
Certain aspects of gauge theories, both at classical and at quantum
level, seem presently most suitable treated in terms of the so-called
antibracket-antifield formalism \cite{bv81} (see \cite{rev} for recent
reviews). As currently formulated, this proposal
relies on BRST invariance as fundamental principle and the use of sources
to deal with BRST transformations \cite{brst} \cite{Zinn}. In this way,
when quantum aspects are under consideration, the field-antifield
formalism (in short, FA) resembles the BRST approach since, in fact,
the sources for BRST transformations of the latter are nothing but the
antifields of the former. The Batalin-Vilkovisky approach, however, not
only encompasses these previous ideas based on BRST invariance
for quantizing gauge theories but also extends and generalizes them to
more complicated situations (open algebras, reducible systems, etc.).

At the classical level, the FA formalism gives a general recipe
to construct, out of a classical gauge action $S_0(\phi)$
and its gauge structure,
a gauge-fixed action ${\gh S}(\Phi^A)$,
suitable for path integral quantization,
its BRST symmetry, $\delta\Phi^A=R^A(\Phi)$,
and the higher order structure functions, $R^{A_n\ldots A_1}(\Phi)$,
characterizing the underlying structure of the classical BRST symmetry.
In practise, this is accomplished by first constructing
from $S_0(\phi)$ and its gauge transformations,
an action $S(z)$ in the so-called classical basis \ct{vp94} of fields and
antifields $z^a=\{\Phi^A,\Phi^*_A\}$, $A=1,\ldots,N$, $a=1,\ldots,2N$,
subject to the boundary conditions:
\begin{enumerate}
\item
Classical limit: $\restric{S(\Phi,\Phi^*)}{\Phi^*=0}= S_0(\phi)$,
\item
Properness condition: ${\rm rank}\restric{(S_{ab})}{\rm on-shell}=N,$
with $S_{ab}\equiv
\left(\frac{\partial_l\partial_r S}{\partial z^a\partial z^b}\right),$
and where on-shell means on the
surface $\left\{\frac{\partial_r S}{\partial z^a}=0\right\}$;
\end{enumerate}
and satisfying the classical master equation
\be
   (S,S)=0,
\label{cme}
\ee
defined in terms of an odd symplectic structure, $(\cdot,\cdot)$, called
antibracket
\be
 (X,Y) = \frac{\partial_r X}{\partial z^a } \zeta^{ab}
         \frac{\partial_l Y}{\partial z^b }
\ , \quad \quad {\rm where } \quad
  \zeta^{ab} \equiv (z^a, z^b)=
  \left(
  \begin{array}{cc}
   0 & \delta^A_B\\
   -\delta^A_B & 0
  \end{array}\right).
\label{antibracket}
\ee

Afterwards,
by further performing a canonical transformation in the antibracket
sense from the classical basis to the so-called gauge-fixed basis
\ct{bv81,vp94}, the BRST structure functions appear to
be the coefficients in the antifield (or source) expansion of $S$
\bea
  S(\Phi,\Phi^*)&=&\gh S(\Phi)+\Phi^*_A R^A(\Phi)
  +\frac12\Phi^*_A\Phi^*_B R^{BA}(\Phi)+\ldots
\nonumber\\
  &&+\frac1{n!}\Phi^*_{A_1}\ldots\Phi^*_{A_n} R^{A_n\ldots A_1}(\Phi)
  +\ldots
\label{gfps}
\eea
while the properness requirement translates in this basis to the condition
$$
  {\rm rank}\restric{(\gh S_{AB})}{\rm on-shell}=N,
   \quad\quad\mbox{with}\quad\quad
   \gh S_{AB}\equiv
   \left(\frac{\partial_l\partial_r \gh S(\Phi)}
   {\partial \Phi^A\partial \Phi^B}\right ),
$$
i.e., propagators are well defined and the usual perturbation theory can
be developped. Under such conditions, all the relations between the
structure functions which characterize algebraically the classical BRST
symmetry are completely encoded in the set of equations coming from
\bref{cme}, which may equivalently be called the classical BRST Ward
identity.

Quantum corrections to this classical BRST symmetry and its underlying
structure are most suitable analyzed in terms
of the quantum counterpart of $S$ \bref{gfps}, i.e.,
by considering the effective action $\Gamma(\Phi, \Phi^*)$ constructed,
via the usual Legendre transformation with respect the sources $J_A$,
from the generating functional
\be
  Z(J,\Phi^*)=\int\gh D \Phi\exp\left\{\frac{i}{\hbar}
  [W(\Phi,\Phi^*)+J_A\Phi^A]\right\},
\label{generating funct}
\ee
where the quantum action $W$
\be
   W= S+\sum^\infty_{p=1}{\hbar}^p M_p,
\label{quantum action}
\ee
anticipates already the presence of local counterterms $M_p$
guaranteeing finiteness of the theory while preserving (as far as
possible) the BRST structure at quantum level.
The quantum BRST structure and its possible breakdown
appear then naturally described by the quantum analog of the
classical BRST Ward identity \bref{cme}
\be
   \frac12(\Gamma,\Gamma)=-i\hbar(\gh A \cdot\Gamma),
\label{ward identity}
\ee
where the obstruction $(\gh A \cdot\Gamma)$ stands for the generating
functional of the 1PI Green functions with one insertion of the composite
field $\gh A$. This composite field $\gh A$
parametrizes thus potential departures from the classical BRST structure
due to quantum corrections and is interpreted as the BRST anomaly.

The standard FA description provides for the anomaly $\gh A$ the expression
\be
   \gh A\equiv
   \left[\Delta W+\frac{i}{2\hbar}(W,W)\right](\Phi,\Phi^*),
\label{anomaly}
\ee
with the operator $\Delta$ defined by
\be
   \Delta\equiv (-1)^{(A+1)}
   \frac{\partial_r}{\partial\Phi^A}
   \frac{\partial_r}{\partial\Phi^*_A},
\label{delta op}
\ee
whereas its $\hbar$ expansion,
$\gh A=\sum_{p=0}\hbar^{p-1} \gh A_{p}$,
yields the form of the $p$-loop BRST anomalies
\bea
     \gh A_0&=&1/2(S,S)\equiv0,
\nonumber\\
     \gh A_1&=&\Delta S+i(M_1, S),
\label{first order}\\
     \gh A_p&=&\Delta M_{p-1}
     +\frac{i}2\sum^{p-1}_{q=1} (M_q,M_{p-q}) +i(M_p, S),
     \quad\quad p\geq 2.
\label{higher order}
\eea
Quantum BRST invariance in this framework is then claimed to be acquired
if the anomaly $\gh A$ \bref{anomaly} in \bref{ward identity} vanishes,
i.e., upon fulfillment through a local object $W$ of the quantum master
equation \cite{bv81}
\be
   \left[\Delta W+\frac{i}{2\hbar}(W,W)\right](\Phi,\Phi^*)=0,
\label{qme}
\ee
which encodes at once the classical master equation
\bref{cme} plus the set of recurrent equations for the counterterms $M_p$
obtained by imposing the vanishing of \bref{first order} and
\bref{higher order}.

The FA quantization program as it stands, however, presents some
drawbacks. First, all the required manipulations
are somewhat formal due to the ill-defined character of the path integral
\bref{generating funct} unless an explicit regularization procedure is
considered. At the level of the BRST Ward identity \bref{ward identity}
this fact manifests itself in the ill-definiteness
of the quantum master equation \bref{qme}, due to the presence of the
operator $\Delta$ \bref{delta op} which generates $\delta(0)$ type
singularities when acting on local functionals as $S$ or $M_p$.
This problematics was first stressed in refs.\,\cite{tnp89,toine1},
where a Pauli-Villars (PV) regularization scheme was
introduced to deal with the quantum master equation \bref{qme} and its
relationship with the existence of anomalies was elucidated%
\footnote{For an updated approach to this subject, see ref.\,\ct{Frank}.}.
The main limitation of this PV regularized version of FA,
however, is its inability to properly regularize
two and higher order loop diagrams, i.e, only first order corrections (one
loop) can be clearly studied.

A second major problem comes from the expression proposed by FA
for the quantities parametrizing two and higher order loop anomalies,
namely \bref{higher order}: while expression \bref{first order}
has been tested in many examples to give the correct one-loop anomaly
\cite{tnp89,vp94,Frank}, it seems clear that, at higher orders,
eq.\,\bref{higher order} can not yield the complete $p$-loop anomaly.
Indeed, on general grounds two and higher loop anomalies get contributions
both from diagrams constructed out of the original action $S$ \bref{gfps}
and, eventually, from new diagrams coming from the finite counterterms
$M_p$, added a posteriori if required by BRST invariance.
The absence in \bref{higher order} of non-trivial contributions coming from
the original action $S$ indicates then that by no means these relations can
generate the complete higher loop anomalies.
In fact, expression \bref{higher order} can be conjectured
to be, as an explicit calculation will further confirm for $p=2$, the
one-loop correction generated by the counterterms $M_k$, $k<p$, to the
$p$-loop anomaly. In view of that, the quantum master equation \bref{qme}
in its present form can only be considered at most as the ``one-loop
form'' of a more general expression, possibly defined in terms of suitable
quantum generalizations of the operator $\Delta$ \bref{delta op},
$\Delta_q=\Delta+\hbar\Delta_1+\ldots$, and of
the antibracket \bref{antibracket} involved in its definition, whose
closed form is still not known. This indicates that fundamental pieces,
the ones reproducing precisely two and higher order loop anomalies, are
missed in standard FA derivations of the quantum master equation
\bref{qme}, so that the proposed FA form \bref{anomaly} for the BRST
anomaly is incomplete at two and higher loops.
It should be stressed, however, that this incompleteness by no
means questions the generic form of the BRST Ward identity
\bref{ward identity} --ensured, as pointed out for instance in \ct{hlw90},
by the celebrated Lam's action principle \ct{Lam72}-- but {\it only the
specific form} \bref{anomaly} {\it of the anomaly insertion proposed in the
standard FA formalism and its formal derivation.}

Confrontation with these drawbacks, thus, leads to looking for
a regularized version of the FA formalism being able to
deal with higher order loop corrections
and to describe the correct structure of the higher
order contributions to the ``one-loop'' BRST anomaly \bref{anomaly}.
In this paper, we pursue the first part of this program by reconsidering
the ideas of a new regularization method, the so-called nonlocal
regularization, recently introduced in \ct{emkw91,kw92,kw93}.
Basically, this method --originally applied to irreducible
theories with closed gauge algebra-- consists
in modifying the original BRST invariant action
by including nonlocal, higher order interactions depending on a given
cut-off $\Lambda^2$ (nonlocalization), in such a way that the resulting
theory is finite at any order. The regulated theory results then to be
invariant at the classical (tree) level under a nonlocal, distorted version
of the original BRST symmetry. Quantum BRST invariance, in turn,
will require in general further introduction of suitable measure factors,
i.e., the ``nonlocal'' analogs of the counterterms $M_p$, which should also
be regularized along similar lines.

Incorporation of all these ideas in the FA framework appears then to
be natural and fruitful since, while giving the way to extend the
formalism of refs.\,\ct{emkw91,kw92,kw93} to general gauge theories (open
algebras, reducible systems, etc), it also provides as a result a fully
regularized version of the antibracket-antifield formalism, suitable for
the study of
the BRST Ward identity \bref{ward identity} proposed in this framework and
related anomaly issues. In the end, the resulting nonlocally regularized
FA formalism is seen to properly deal with higher order loop corrections
and anomalies when computing them directly from the BRST variation of the
effective action $\Gamma$ --as the considered example, $W_3$, shows.
However, and as stated before, when approached by using the form
\bref{anomaly} of the anomaly, it only reproduces correctly one-loop anomalies
and one-loop corrections to higher loop anomalies coming from
counterterms, providing thus an explicit verification of the previous
assertion. The analysis of the complete form of the anomaly
and of the quantum master equation, however, lies outside the scope
of this paper and its study is left as an open problem.

The paper has been organized as follows.
In section 2, the nonlocal regularization method of refs.\ct{kw92,kw93}
is briefly reviewed in order to set out notation and to slightly
improve it to describe in generality systems with either first or second
order kinetic term in space-time derivatives. Section 3 deals with the
extension of the previous method to general gauge theories by
reformulating it according to the
ideas of the antibracket-antifield formalism, both at classical and at
quantum level. In particular, a prescription is given to obtain the
regularized form of the proper solution of the classical master equation
$S$ and of its quantum extension $W$ and a regularized
version of the quantum master equation is provided after analyzing
the action of $\Delta$ on this regularized quantum action.
The theoretical framework is then exemplified in section 4
by applying it to chiral $W_3$ gravity. First of all, the one and two-loop
orders of the regularized quantum master equation are used to
compute the complete one-loop anomaly and the correction to the universal
two-loop anomaly produced by the addition of a one-loop counterterm,
respectively, while the universal two-loop anomaly is afterwards evaluated
from the BRST variation of the nonlocally regulated effective action.
Section 5 summarizes the conclusions and indicates where the
incompleteness of the quantum master equation can originate by analyzing
its standard FA derivation. Finally, an appendix devoted to
the computation of the general form of the functional traces involved in
the calculations of section 4 is provided.

\section{Nonlocal Regularization}

\hspace{\parindent}%
The aim of nonlocal regularization is to provide a systematic, field
theoretic formulation of the main ideas contained in Schwinger's proper
time method \ct{schwinger51}. In this approach, propagators
are realized in terms of integrals over real parameters
$t_i\in[0,\infty)$, $i=1,\ldots, n$, so that ultraviolet
divergences become, after integration over loop momenta, singularities
around the region $t_i=0$. Nonlocal regularization proposes to split the
original divergent loop integrals as a sum over loop contributions coming
from the original fields with modified propagators plus extra loop
contributions generated by a set of auxiliary fields, in such a way that
original singularities be contained in the loops composed solely of
auxiliary fields. Elimination of these auxiliary fields by putting them
on--shell gets rid of their quantum fluctuations, and so of its divergent
loops, and regularizes the original theory.

In what follows, we briefly review the nonlocal regularization method along
the lines of refs.\ct{kw92,kw93}, with some slight modifications
in order to describe in a unified way
systems with either first or second order kinetic term in space-time
derivatives. This summary shall also serve to fix our conventions and
notations.

\subsection{The Nonlocally Regulated Action}

\hspace{\parindent}%
Consider a theory defined by a classical action $\gh S(\Phi)$ (as for
instance, the gauge-fixed action in \bref{gfps}), which depends on the set
of fields $\Phi^A$, $A=1,\ldots,N$, with statistics $\eps(\Phi^A)\equiv A$.
The so-called nonlocal regularization applies to theories which have a
sensible perturbative expansion, i.e., to theories for which the
action $\gh S(\Phi)$ admits a decomposition into free and interacting parts
\be
  \gh S(\Phi)= F(\Phi)+ I(\Phi),
  \quad\quad \mbox{\rm with}\quad\quad
   F(\Phi)=\frac12\Phi^A \gh F_{AB} \Phi^B,
\label{original action}
\ee
and where $I(\Phi)$ is assumed to be analytic in $\Phi^A$ around
$\Phi^A=0$.

Introduce now a cut--off or regulating parameter $\Lambda^2$ and
a field independent operator $T_{AB}$, invertible but otherwise arbitrary.
This operator is chosen such that from its inverse $(T^{-1})^{AB}$
and the kinetic operator $\gh F_{AB}$ in \bref{original action}, a second
order derivative ``regulator'' $\gh R^A_{\,B}$ arises through the
combination
$$
   \gh R^A_{\,B}=(T^{-1})^{AC}\gh F_{CB}.
$$
Afterwards, from this object, the smearing operator, $\gh\veps^A_{\,B}$,
and the shadow kinetic operator, $\gh O^{-1}_{AB}$, are constructed
\be
   \gh\veps^A_{\,B}= \exp\left(\frac{\gh R^A_{\,B}}{2\Lambda^2}\right),
\label{smearing op}
\ee
\be
   \gh O^{-1}_{AB}=T_{AC}(\tilde{\gh O}^{-1})^C_{\,B}=
    \left(\frac{\gh F}{\gh\veps^2-1}\right)_{AB},
\label{shadow kinetic op}
\ee
with $(\tilde{\gh O})^A_{\,B}$ defined as
$$
   \tilde{\gh O}^A_{\,B}=
    \left(\frac{\veps^2-1}{\gh R}\right)^A_{\,B}=
   \int^1_0\frac{\dif t}{\Lambda^2}\,
   \exp\left(t\frac{\gh R^A_{\,B}}{\Lambda^2}\right).
$$

Now, for each field $\Phi^A$ introduce an auxiliary field $\Psi^A$ with the
same statistics --the so-called shadow fields-- and couple both sets of
fields through the auxiliary action
\be
  \tilde{\gh S}(\Phi,\Psi)= F(\hat\Phi)-A(\Psi)+  I(\Phi+\Psi),
\label{auxiliary action}
\ee
with $A(\Psi)$, the kinetic term for the auxiliary fields,
constructed with the help of \bref{shadow kinetic op} as
$$
  A(\Psi)=\frac12\Psi^A \gh O^{-1}_{AB} \Psi^B,
$$
and where the ``smeared'' fields $\hat\Phi^A$ appearing in the free part of
the auxiliary action \bref{auxiliary action} are defined, using
\bref{smearing op}, by
\be
  \hat\Phi^A\equiv(\gh\veps^{-1})^A_{\,B}\Phi^B=
  \Phi^B(\gh\veps^{-1})^{\,A}_B,
\label{smeared field}
\ee
with $(\gh\veps^{-1})^{\,A}_B$ the supertranspose of
$(\gh\veps^{-1})^A_{\,B}$, i.e.,
$(\gh\veps^{-1})^{\,A}_B=(\gh\veps^{-1})^A_{\,B} (-1)^{B(A+B)}$.

{}From the above construction, it should be clear that the perturbative
theory described by \bref{auxiliary action}, when only external $\Phi$
lines are considered, and \bref{original action} are exactly the same.
Indeed, for a given diagram, $\Phi$ external lines connect either to a
smeared propagator
\be
   \left(\frac{i\gh\veps^2}{\gh F+i\eps}\right)^{AB}=
   -i \left[\int^\infty_1\frac{\dif t}{\Lambda^2}\,
   \exp\left(t\frac{\gh R^A_{\,C}}{\Lambda^2}\right)\right](T^{-1})^{CB},
\label{smeared propagator}
\ee
or to a shadow propagator
\be
   -i \gh O^{AB}= \left(\frac{i(1-\gh\veps^2)}{\gh F}\right)^{AB}=
   -i \left[\int^1_0\frac{\dif t}{\Lambda^2}\,
   \exp\left(t\frac{\gh R^A_{\,C}}{\Lambda^2}\right)\right](T^{-1})^{CB},
\label{shadow propagator}
\ee
which diagramatically are going to be represented by ``unbarred'' and
``barred'' lines, respectively, (Fig.\,1)

\begin{picture}(40000,8000)


\drawline\fermion[\E\REG](8000,4000)[8000]


\drawline\fermion[\E\REG](27000,4000)[8000]
\global\advance\pmidx by -100
\put(\pmidx,3500){$\rule{1mm}{3mm}$}
\global\advance\pmidx by 100

\end{picture}

\centerline{Fig.\,1. Unbarred and barred propagators.}

\vskip 0.7cm
\noindent
On the other hand, the specific form of
the interaction in \bref{auxiliary action}, $I(\Phi+\Psi)$,
associates, to a given original diagram with $n$ internal lines,
$2^n$ diagrams consisting in all possible combinations formed with
both smeared \bref{smeared propagator}
and shadow \bref{shadow propagator} internal propagators.
Then, the sum of \bref{smeared propagator} and \bref{shadow propagator}
being the original $\Phi$ propagator, it is concluded that the
sum of all these $2^n$ contributions yields the original
diagram with $n$ original propagators in its internal lines. However,
the special form of the shadow propagator \bref{shadow propagator}
as an integral over $t$ in the region $[0,1]$ leads the
contribution composed solely from ``barred'' lines to isolate
the divergent part of the original diagram, i.e., the one coming from the
integration over the hypercube $[0,1]^n$ in parameter space. Cancellation
of this diagram contribution and, in the general case, of the quantum
fluctuations of the shadow fields, regularizes thus the theory.

The process of eliminating quantum fluctuations associated uniquely
with shadow fields, or conversely, of the closed loops formed solely
with barred lines by hand,
is equivalently implemented at the path integral level
by putting the auxiliary fields $\Psi$ on-shell. In this second strategy,
the classical shadow field equations of motion
\be
  \frac{\partial_r \tilde{\gh S}(\Phi,\Psi)}{\partial \Psi^A}=0
  \quad\Rightarrow\quad
  \Psi^A=\left(\rder{I}{\Phi^B}(\Phi+\Psi)\right)\gh O^{BA},
\label{shadow eqs motion}
\ee
should be solved, in general, in a perturbative fashion
and its classical solution $\bar\Psi_0(\Phi)$ substituted in the
auxiliary action \bref{auxiliary action}. The result of this process is the
nonlocalized action to be used in regularized computations
\be
  \gh S_{\Lambda}(\Phi)\equiv \tilde{\gh S}(\Phi,\bar\Psi_0(\Phi)).
\label{nonlocal action}
\ee

The expansion of $\gh S_{\Lambda}(\Phi)$ in $\bar\Psi_0$ can then be seen
to generate the smeared kinetic term $F(\hat\Phi)$, the original
interaction $I(\Phi)$ plus an infinite series of new nonlocal interaction
terms, whose effect is completely equivalent to that of
the mixed loops of unbarred and barred lines in the previous
formulation, and which arise from them after putting the shadow fields
on-shell. All these new nonlocal interaction terms
are of $O(\frac1{\Lambda^2})$ and, therefore, vanishing in the limit
$\Lambda^2\rightarrow\infty$, so that
$\gh S_{\Lambda}(\Phi)\rightarrow\gh S(\Phi)$ in this limit and
the original theory is recovered. This result can also be obtained by
directly considering in $\gh S_{\Lambda}(\Phi)$ the following limits
when $\Lambda^2\rightarrow\infty$:
\be
  \veps\rightarrow 1,\quad\quad
  \gh O\rightarrow 0,\quad\quad
  \bar\Psi_0(\Phi)\rightarrow 0.
\label{limits}
\ee

In conclusion, the nonlocally regulated theory can thus be realized in
two ways: either by using the auxiliary action \bref{auxiliary action} and
eliminating by hand closed loops formed solely with barred lines, or by
putting the fields $\Psi$ on--shell. The first strategy is the best
when performing diagramatic calculations, since the Feynman
rules which correspond to the auxiliary action \bref{auxiliary action} are
essentially as simple as those of the original local theory. Use of the
nonlocally regulated action \bref{nonlocal action}, instead,
is more convenient when doing algebraic manipulations, as the ones
involved in the analysis of the regulated version of the quantum master
equation \bref{qme}. Examples of both procedures will be explicitly
considered in the specific model of chiral $W_3$ gravity.

\subsection{Nonlocally Regulated Symmetries}

\hspace{\parindent}%
Any local quantum field theory can be made ultraviolet finite just by
introducing the smearing operator \bref{smearing op} and by working with
the modified action $F(\hat\Phi)+ I(\Phi)$, $\hat\Phi$ being the smeared
field \bref{smeared field} \ct{emkw91,kw92,kw93}%
\footnote{See also \ct{polchinsky84} for an earlier approach
and \ct{ms94} for an alternative formulation of this idea in the context
of chiral perturbation theory.}.
However, this form of nonlocalization generally spoils any sort of
gauge symmetry or its associated BRST symmetry, leading to the breakdown of
the corresponding Ward identities already at the tree level
and threatening all the benefits derived by its use in the study of
perturbative unitarity and renormalizability issues.

The form of nonlocalization presented above, instead, has the merit
of preserving at tree level a distorted version of any of the original
continuous symmetries of the theory, and in particular, of the BRST (or
of the original gauge) symmetry. Indeed, assume that the original action
\bref{original action} is invariant under the infinitesimal
transformation
\be
   \delta\Phi^A= R^A(\Phi).
\label{original sym}
\ee
Then, the auxiliary action results to be invariant under the
auxiliary infinitesimal transformations
\be
   \tilde\delta\Phi^A= (\gh\veps^2)^A_{\,B} R^B(\Phi+\Psi),
   \quad\quad
   \tilde\delta\Psi^A= (1-\gh\veps^2)^A_{\,B} R^B(\Phi+\Psi),
\label{auxiliary transf}
\ee
while the nonlocally regulated action $\gh S_{\Lambda}(\Phi)$
\bref{nonlocal action} becomes invariant under
\be
   \delta_{\Lambda}\Phi^A= (\gh\veps^2)^A_{\,B} R^B(\Phi+\bar\Psi_0(\Phi)),
\label{nonlocal transf}
\ee
with $\bar\Psi_0(\Phi)$ the classical solution of
\bref{shadow eqs motion}. The proof of these statements is
straightforward and can be found in the original reference \ct{kw92}, to
which the reader is referred for further details.

The nonlocalization procedure provides thus a definite way to
regularize the continuous symmetries of a given theory. However, although
apparently symmetries are maintained, its underlying structure is changed
somehow. Indeed, it is possible to see \ct{kw92} that, for instance, if the
original symmetry \bref{original sym} has a closed algebra, its
corresponding
nonlocal version $\delta_{\Lambda}$ \bref{nonlocal transf} closes in
general only on-shell. In other words, if, simbolically
$[\delta_1, \delta_2]=\delta_3$, then in general
\be
   [\delta_{\Lambda,1}, \delta_{\Lambda,2}]\Phi^A=
   \delta_{\Lambda,3}\Phi^A+
   \left[(\gh\veps^2)^A_{\,B}\,\Omega^{BC}_{12}(\Phi+\bar\Psi_0(\Phi))\,
   (\gh\veps^2)^{\,D}_C (-1)^D\right]
   \rder{\gh S_{\Lambda}(\Phi)}{\Phi^D},
\label{original nonclosure}
\ee
with $\Omega^{AB}_{12}$ given by
$$
   \Omega^{AB}_{12}(\Phi)=
   \left(\rder{R^A_1}{\Phi^C} \, K^{CD} \,
   \lder{R^B_2}{\Phi^D}-(1\leftrightarrow 2) \right)(\Phi),
$$
and where the (inverse of the) operator $K^{AB}$ is defined as
\be
    (K^{-1})_{AB}(\Phi)=
    \left[(\gh O^{-1})_{AB}- \frac{\partial_l\partial_r I(\Phi)}
    {\partial\Phi^A\partial\Phi^B}\right].
\label{k operator}
\ee
This phenomenon is well-known to happen upon elimination of a given
subset of (auxiliary) fields through their classical equations of motion
\cite{rev,henn}.

Continuing in this way, that is to say, by taking more and more
commutators of nonlocally regulated symmetries \bref{nonlocal transf}, the
higher order structure functions characterizing the original symmetry are
seen to be modified in much the same direction by the nonlocalization
process. In summary, the complete symmetry structure associated with the
nonlocally regulated transformation \bref{nonlocal transf} results to be a
distorted version of the original one: not only the action and the
transformations are distorted, but also the underlying structure functions
characterizing the original symmetry are also regulated and changed.
The nonlocal regularization procedure works therefore by distorting or
regularizing the complete classical structure of the original theory,
in such a way that in the limit $\Lambda^2\rightarrow\infty$ the regulated
structure converges to the original one.

\section{Nonlocally regularized Antibracket--Antifield formalism}

\hspace{\parindent}%
In view of the previous results, the question naturally arises,
for a given BRST quantized gauge theory, of the relationship between the
original structure functions and its underlying algebraic BRST structure
with their regulated versions. It is here where the antibracket-antifield
framework turns out to be useful in characterizing and determining the
structure of the regulated BRST symmetry in terms of
the original one. The results in ref.\,\ct{kw92},
mainly designed for irreducible theories with closed gauge algebra,
partially answered the question by giving the explicit
form of the regulated action \bref{nonlocal action}
and of the BRST transformations \bref{nonlocal transf}. Using these forms
as starting point, or as boundary conditions, and combining them with
well-known FA results, the regulated BRST classical structure of a general
gauge theory can be completely elucidated from the
knowledge of the original one. As a byproduct of this nonlocalization
process, and after extending it to the quantum action $W$, a nonlocally
regularized antibracket-antifield formalism comes out in a natural way.

\subsection{Nonlocalization of the proper solution}

\hspace{\parindent}%
Consider a FA quantized gauge theory and assume that the gauge
fixed action $\gh S(\Phi)$ in the proper solution of the classical master
equation \bref{gfps} admits a suitable perturbative decomposition
of the form \bref{original action}.
{}From the general results stated in the introduction, it is concluded that
the regulated version of the structure functions should come out as
coefficients in the antifield expansion of a proper solution having as
lowest order terms the nonlocal action
\bref{nonlocal action} and its nonlocal BRST transformations
\bref{nonlocal transf}. The uniqueness of these
regulated structure functions should be understood, of course, modulo
canonical transformations.

This nonlocal proper solution can be obtained by rewriting the
process described in sect.\,2 according to FA ideas.
Enlarge thus first the original field-antifield space by introducing
the shadow fields and their antifields $\{\Psi^A,\Psi^*_A\}$.
Then, in this extended space,
an auxiliary proper solution should be constructed as a preliminary step,
which incorporates
the auxiliary action $\tilde{\gh S}(\Phi,\Psi)$ \bref{auxiliary action},
corresponding to the gauge-fixed action $\gh S(\Phi)$,
its BRST symmetry \bref{auxiliary transf}
and the yet unknown associated higher order structure functions.
In doing so, it is realized that the auxiliary BRST transformations
\bref{auxiliary transf}, which should appear in the auxiliary proper
solution as the first order term in the antifields
$$
  \left[\Phi^*_A(\gh\veps^2)^A_{\,B}+
  \Psi^*_A (1-\gh\veps^2)^A_{\,B}\right]
  R^B(\Phi+\Psi),
$$
can be obtained from the term $\Phi^*_A R^A(\Phi)$ in the original proper
solution through the replacements
\bea
  &&\Phi^*_A\rightarrow\left[\Phi^*_B(\gh\veps^2)^B_{\,A}+
  \Psi^*_B (1-\gh\veps^2)^B_{\,A}\right]\equiv\Theta^*_A,
\label{replacement 1}\\
  &&R^A(\Phi)\rightarrow R^A(\Phi+\Psi)\equiv R^A(\Theta).
\label{replacement 2}
\eea
It is therefore tempting to extend this rule to the higher order
antifield terms in \bref{gfps} in order to reconstruct the auxiliary higher
order structure functions, i.e., to consider \bref{replacement 1} and
\be
   R^{A_n\ldots A_1}(\Phi)\rightarrow R^{A_n\ldots A_1}(\Phi+\Psi)=
   R^{A_n\ldots A_1}(\Theta),
\label{replacement 3}
\ee
and write as an ansatz for the auxiliary proper solution
\bea
  \tilde{S}(\Phi,\Phi^*;\Psi, \Psi^*)&=&
  \tilde{\gh S}(\Phi,\Psi)+\Theta^*_A R^A(\Theta)+
  \frac12 \Theta^*_A\Theta^*_B R^{BA}(\Theta)+\ldots
\nonumber\\
  &&+\frac1{n!}\Theta^*_{A_1}\ldots\Theta^*_{A_n}
  R^{A_n\ldots A_1}(\Theta)+\ldots.
\label{auxiliary proper}
\eea

It is not difficult to check that \bref{auxiliary proper} is indeed a proper
solution of \bref{cme} in the extended space. One should realize, first of
all, that the linear combinations $\{\Theta^A, \Theta^*_A\}$ defined in
\bref{replacement 1}, \bref{replacement 2},
from which the antifield dependent part of \bref{auxiliary proper} is
constructed, are conjugated variables, i.e.,
$(\Theta^A,\Theta^*_B)=\delta^A_{\,B}$. This fact suggests to
use a new set of fields and antifields
obtained by completing the subset $\{\Theta^A,\Theta^*_A\}$
with the linear combinations $\{\Sigma^A,\Sigma^*_A\}$ defined as
$$
  \Sigma^A=
  \left[(1-\gh\veps^2)^A_{\,B}\Phi^B-(\gh\veps^2)^A_{\,B}\Psi^B\right],
  \quad\quad
  \Sigma^*_A= \Phi^*_A-\Psi^*_A,
$$
in such a way that the linear transformation
\be
  \{\Phi^A,\Phi^*_A;\Psi^A,\Psi^*_A\}\rightarrow
   \{\Theta^A,\Theta^*_A;\Sigma^A,\Sigma^*_A\},
\label{canonical transf}
\ee
be canonical in the antibracket sense.
In terms of this new set of variables the auxiliary action
\bref{auxiliary proper} acquires the form
\be
  \tilde{S}(\Theta,\Theta^*;\Sigma, \Sigma^*)=
  S(\Theta,\Theta^*)+
  \frac12\Sigma^A
  \left[\frac{\gh F}{\veps^2}+ \frac{\gh F}{(1-\veps^2)}\right]_{AB}
  \Sigma^B,
\label{canonical transf aux}
\ee
where $S(\Theta,\Theta^*)$ is the original proper solution
\bref{gfps} with arguments $\{\Theta^A,\Theta^*_A\}$.
Therefore, since no antifields $\Sigma^*$ are present and
$S(\Theta,\Theta^*)$ verifies the classical master equation,
the action \bref{canonical transf aux} fulfills \bref{cme} as well.
The canonical nature of the linear transformation
\bref{canonical transf} further ensures the fulfillment of \bref{cme}
by the auxiliary action \bref{auxiliary proper}. On the other hand,
properness of \bref{canonical transf aux}, derived from the
properness of the original
solution $S$ and the fact that the kinetic term of the
$\Sigma^A$ fields has maximum rank, guarantees properness of
the auxiliary action \bref{auxiliary proper} due to the conservation of the
proper character of a solution of \bref{cme} upon canonical
transformations. It can also be checked by direct inspection of the
auxiliary action \bref{auxiliary proper}.
This completes the first part of the process, thus indicating that
replacements \bref{replacement 1}, \bref{replacement 2} and
\bref{replacement 3} provide an effective way to describe the structure
functions characterizing the auxiliary BRST structure in terms of the
original BRST structure, without performing any concrete computation.

The next step of the process, namely, nonlocalization, is acquired by
eliminating the shadow fields and antifields by means of the standard
elimination process of auxiliary fields in the FA framework%
\footnote{For an study of the elimination process of auxiliary fields and
antifields in the antibracket-antifield formalism and its consequences, the
reader is referred to \cite{rev,henn}.},
that is, by substituting the shadow fields by the solutions of their
classical equations of motion, while putting their antifields to zero.
In other words, the candidate to nonlocal proper solution
is obtained from \bref{auxiliary proper} as
\be
   S_{\Lambda}(\Phi,\Phi^*)=\tilde{S}(\Phi,\Phi^*;\bar\Psi, \Psi^*=0),
\label{nonlocal proper}
\ee
with $\bar\Psi\equiv\bar\Psi(\Phi,\Phi^*)$ the solution of the classical
equations of motion
\be
   \rder{\tilde{S}(\Phi,\Phi^*;\Psi, \Psi^*=0)}{\Psi^A}=0.
\label{eq motion}
\ee
The substitution of the subset of fields
$\{\Psi^A, \Psi^*_A\}$ in this way leads \bref{nonlocal proper} to fulfill
again \bref{cme}. Indeed, a simple computation using relations
\be
   \rder{S_{\Lambda}}{\Phi^A}= \restric{\rder{\tilde S}{\Phi^A}}{},
   \quad\quad
   \lder{S_{\Lambda}}{\Phi^*_A}= \restric{\lder{\tilde S}{\Phi^*_A}}{},
\label{relations der}
\ee
where the above restriction means on the surface
$\{\Psi=\bar\Psi(\Phi,\Phi^*), \Psi^*=0\}$, and
which hold as a consequence of \bref{eq motion}, leads to
\be
  \frac12(S_{\Lambda}, S_{\Lambda})=
  \restric{\rder{\tilde S}{\Phi^A}\lder{\tilde S}{\Phi^*_A}}{}=
  \restric{\left(\rder{\tilde S}{\Phi^A}\lder{\tilde S}{\Phi^*_A}+
  \rder{\tilde S}{\Psi^A}\lder{\tilde S}{\Psi^*_A}\right)}{}=
  \frac12\restric{(\tilde S, \tilde S)}{}=0.
\label{s lambda 0}
\ee

On the other hand, in order to verify properness of
\bref{nonlocal proper}, as well as to finally determine the nonlocally
regulated version of the structure coefficients, it is worth to
investigate in more detail the antifield expansion of $S_\Lambda$.
To this end, the equations of motion \bref{eq motion} for the fields
$\Psi^A$, which explicitly read
\be
  \Psi^A=
  \left[\rder{I}{\Phi^B}(\Phi+\Psi)+
  \Phi^*_C(\gh\veps^2)^C_{\,D} R^D_{\,B}(\Phi+\Psi)+\gh O((\Phi^*)^2)
  \right]\gh O^{BA},
\label{eq motion 2}
\ee
with $R^A_{\,B}=\rder{R^A(\Phi)}{\Phi^B}$, can be solved perturbatively in
powers of antifields in the standard way. The lowest order of equation
\bref{eq motion 2}
\be
  \Psi^A=
  \left(\rder{I}{\Phi^B}(\Phi+\Psi)\right)\gh O^{BA},
\label{shadow eqs motion 0}
\ee
results to be equation \bref{shadow eqs motion}, as expected. Expanding
around its solution $\bar\Psi_0$ at first order in the antifields,
$\Psi^A_1(\Phi,\Phi^*)=\bar\Psi^A_0(\Phi) +\Phi^*_B D^{BA}(\Phi)$, and
plugging this expression in \bref{eq motion 2}, yields
\be
  \bar\Psi^A_1(\Phi,\Phi^*)=\bar\Psi^A_0(\Phi) +
  \Phi^*_B (\gh\veps^2)^B_{\,C}R^C_{\,D} K^{DA}(\Phi+\bar\Psi_0),
\label{shadow eqs motion 1}
\ee
with (the inverse of) $K^{DA}(\Phi)$ given by
\bref{k operator}. Applying recursively this procedure,
an expression can be obtained for $\bar\Psi(\Phi,\Phi^*)$ at any desired
order in antifields in terms of the antifield independent solution
$\bar\Psi_0$ of eq.\,\bref{shadow eqs motion 0}.

Solution \bref{shadow eqs motion 1} is enough to figure out the form of the
regulated structure functions up to second order in the antifields and to
compare the results with the ones presented in sect.\,2. Indeed,
recall the form of the nonlocal proper solution \bref{nonlocal proper} in
terms of the auxiliary proper solution \bref{auxiliary proper}.
The lowest order term of \bref{auxiliary proper} yields
$$
  \tilde{\gh S}(\Phi,\bar\Psi_1)=
  \tilde{\gh S}(\Phi,\bar\Psi_0)-
  \frac12
  \left[\Phi^*_A\, (\gh\veps^2)^A_{\,C}\,R^C_{\,D}\,
  \Phi^*_B\, (\gh\veps^2)^B_{\,F}\,R^F_{\,E}\,
   K^{ED}\right](\Phi+\bar\Psi_0),
$$
the first order term results in
$$
  \Phi^*_A(\gh\veps^2)^A_{\,B} R^B(\Phi+\bar\Psi_1)=
$$
$$
  \Phi^*_A(\gh\veps^2)^A_{\,B} R^B(\Phi+\bar\Psi_0)+
  \left[\Phi^*_A\, (\gh\veps^2)^A_{\,C}\,R^C_{\,D}\,
  \Phi^*_B\, (\gh\veps^2)^B_{\,F}\,R^F_{\,E}\,
  K^{ED}\right](\Phi+\bar\Psi_0),
$$
while the second order term gives the same expression evaluated in
$\bar\Psi_0$ plus corrections of $O((\Phi^*)^3)$. Collecting all
these expressions, the complete form of the nonlocal proper solution arises
up to second order in the antifields
\bea
   S_{\Lambda}(\Phi,\Phi^*)&=&
   \tilde{\gh S}(\Phi,\bar\Psi_0)+
   \Phi^*_A(\gh\veps^2)^A_{\,B} R^B(\Phi+\bar\Psi_0)
\nonumber\\
   &&+\frac12 \Phi^*_A\, (\gh\veps^2)^A_{\,C}\,
  \Phi^*_B\, (\gh\veps^2)^B_{\,D}\,
  \left[R^{DC}+ R^D_{\,F} K^{FE} R_E^{\,C}\right](\Phi+\bar\Psi_0)
  +\ldots,
\label{final nlocal proper}
\eea
with $R_B^{\,A}= \lder{R^A(\Phi)}{\Phi^B}$.

The lowest order antifield terms in \bref{final nlocal proper} are
precisely the (gauge-fixed) nonlocal action \bref{nonlocal action} and
its nonlocal BRST transformations \bref{nonlocal transf}, as required.
Properness of $S_\Lambda$ comes now from the fact that the new interaction
terms, of $O(\frac1{\Lambda^2})$, appearing in the
(gauge-fixed) nonlocal action
\bref{nonlocal action} can not alter the maximum rank of the hessian
of the original gauge-fixed action \bref{original action}.
Continuing in this way, the quadratic piece in antifields provides the
regulated form of the on-shell nilpotency structure functions, whose form
could also have been inferred from computation
\bref{original nonclosure} when dealing with an open algebra. In
conclusion, up to this point the results of sect.\,2 are completely
reproduced in the FA framework.
Further determination of the regulated form of the higher order structure
coefficients would proceed in the same fashion, that is,
by solving \bref{eq motion} at second and higher order in antifields and
plugging the solution in \bref{nonlocal proper}.

In summary, expression \bref{nonlocal proper} supplemented with
\bref{eq motion} encodes in a compact way all the information about
the structure of the nonlocally regulated BRST symmetry and
naturally provides a nonlocally regularized proper solution to be used in
algebraic perturbative computations. As in the original formulation,
in the limit $\Lambda^2\rightarrow\infty$, the nonlocally regulated
gauge-fixed action $\tilde{\gh S}(\Phi,\bar\Psi_0)$ becomes the original
one $\gh S(\Phi)$, while the antifield dependent part of
\bref{final nlocal proper} acquires the original form as well, as a
consequence of the limits \bref{limits}. In the unregulated limit thus
\bref{final nlocal proper} converges to the original proper solution
\bref{gfps} and the original local theory is recovered.
This concludes the characterization of the regulated
antibracket-antifield formalism at classical level.

\subsection{Nonlocally regulated quantum master equation}

\hspace{\parindent}%
{}From the interplay between the nonlocal
regularization method and the antibracket-antifield formalism,
a nonlocally regularized proper solution $S_\Lambda$ of the classical
master equation arises. In this sense, the classical BRST structure can
be considered to be preserved at regularized level. Quantization of the
theory, however, as indicated by the BRST Ward identity
\bref{ward identity}, requires in general the addition of extra
counterterms or interactions $M_p$ in order to preserve the
quantum counterpart of the classical BRST structure, i.e., the
substitution of the classical action $S$ \bref{gfps} by a suitable
quantum action $W$ \bref{quantum action}. This fact was already stressed
in the original references \ct{emkw91,kw92,kw93}, although that approach
seemed to suggest that only one-loop corrections $M_1$ would eventually
suffice to acquire BRST invariance. Instead, the FA formalism indicates
that generally two and higher order loop corrections should also be
considered. The question arises then of how regularization of these new
interaction terms, or conversely, of the new theory described by $W$,
should proceed.

The way in which nonlocalization acts on the free part $F(\Phi)$ and on the
complete interaction part $\gh I(\Phi,\Phi^*)$ of the original proper
solution, defined as
$$
  \gh I(\Phi,\Phi^*)\equiv I(\Phi) +\Phi^*_A R^A(\Phi)
  +\frac12\Phi^*_A\Phi^*_B R^{BA}(\Phi)+\ldots,
$$
provides also a definite way to regularize interactions coming from
counterterms $M_p$. Indeed,
the process described in the previous section can be summarized by first
constructing the auxiliary free and interaction parts
$$
   \tilde F(\Phi,\Psi)=F(\hat\Phi)-A(\Psi),\quad\quad
   \tilde{\gh I}(\Phi,\Phi^*,\Psi,\Psi^*)= \gh I(\Theta,\Theta^*),
$$
with $\{\Theta,\Theta^*\}$ the linear combinations defined in
\bref{replacement 1}, \bref{replacement 2}, and by putting afterwards
the auxiliary fields $\Psi$ on-shell and its antifields to zero
$$
   F_\Lambda(\Phi,\Phi^*)= \tilde F(\Phi,\bar\Psi_0),\quad\quad
   \gh I_\Lambda(\Phi,\Phi^*)=
   \tilde{\gh I}(\Phi+\bar\Psi_0, \Phi^*\gh\veps^2),
$$
so that in the end $S_\Lambda=F_\Lambda+\gh I_\Lambda$.

The nonlocalization procedure for the quantum action $W$
\bref{quantum action}, expressed as
\be
   W= F+\gh I+\sum^\infty_{p=1}{\hbar}^p M_p\equiv F+\gh Y,
\label{quantum action 2}
\ee
which defines the generalized ``quantum'' interaction $\gh Y$,
should then proceed along the same lines, i.e., constructing
first an auxiliary quantum action $\tilde W$
\bea
  \tilde W(\Phi,\Phi^*,\Psi,\Psi^*)&=&
  F(\hat\Phi)-A(\Psi)+ \gh Y(\Theta,\Theta^*)
\nonumber\\
  &=& W(\Theta,\Theta^*)+ \frac12\Sigma^A
  \left[\frac{\gh F}{\veps^2}+ \frac{\gh F}{(1-\veps^2)}\right]_{AB}
  \Sigma^B,
\label{aux quantum action}
\eea
and eliminating afterwards the auxiliary fields and antifields in the usual
way, i.e.,
\be
  W_\Lambda(\Phi,\Phi^*)= \tilde W(\Phi,\Phi^*,\bar\Psi_q,\Psi^*=0),
\label{w lambda}
\ee
where the quantum on-shell auxiliary fields $\bar\Psi_q$ are now the
solutions of the equations of motion associated with the auxiliary
quantum action $\tilde W$ \bref{aux quantum action}
\be
   \rder{\tilde{W}(\Phi,\Phi^*;\Psi, \Psi^*=0)}{\Psi^A}=0
  \quad\Rightarrow\quad
  \Psi^A=\left(\rder{\gh Y}{\Phi^B}(\Phi+\Psi,\Phi^*\gh\veps^2)
  \right)\gh O^{BA}.
\label{q equations motion}
\ee

The quantum on-shell auxiliary fields $\bar\Psi_q$ solving
\bref{q equations motion} are seen afterwards to
contain as lowest order term in $\hbar$ the classical on-shell shadow
fields $\bar\Psi$ verifying \bref{eq motion}. This fact, together with the
form \bref{quantum action} of $W$, indicates that $W_\Lambda$
contains still the nonlocal proper solution $S_\Lambda$
\bref{nonlocal proper} as lowest order term in $\hbar$, while the higher
order terms in $\hbar$ should be interpreted as the
nonlocal form $M_{p,\Lambda}$ of the counterterms $M_p$, i.e.,
$$
  W_\Lambda= S_\Lambda+\sum_{n=1}\hbar^p M_{p,\Lambda}.
$$
Therefore, the nonlocalization procedures for $S$ and $W$ coincide
at tree (classical) level, as expected.

The nonlocally regulated quantum action $W_\Lambda$ \bref{w lambda}
obtained in this way
is thus the object to be used in the path integral \bref{generating funct}
in order to perform regularized perturbative computations.
Defining afterwards $\Gamma_\Lambda$ as the corresponding regulated
effective action and following the same steps used to
derive \bref{ward identity}, which now seems to be meaningful, the
regulated BRST Ward identity comes out
\be
   \frac12(\Gamma_\Lambda,\Gamma_\Lambda)=
   -i\hbar(\gh A_\Lambda \cdot\Gamma_\Lambda),
\label{reg ward ident}
\ee
where the obstruction $\gh A_\Lambda$ acquires now the form
\bref{anomaly} with the substitution $W\rightarrow W_\Lambda$
\be
   \gh A_\Lambda=
   \left[\Delta W_\Lambda+\frac{i}{2\hbar}(W_\Lambda,W_\Lambda)\right].
\label{reg qme}
\ee

Let us now investigate more precisely the form \bref{reg qme} of the
quantum master equation in this regularized framework. Consider first of
all the action of the operator $\Delta$ on $W_\Lambda$
$$
  \Delta W_\Lambda=
 \frac{\partial_r\partial_l W_\Lambda}{\partial\Phi^B\partial\Phi^*_A}
  \delta^B_{\,A}.
$$
Use of the analogs of relations \bref{relations der}
between derivatives of $W_{\Lambda}$ and $\tilde{W}$,
\be
   \rder{W_{\Lambda}}{\Phi^A}= \restric{\rder{\tilde W}{\Phi^A}}{q},
   \quad\quad
   \lder{W_{\Lambda}}{\Phi^*_A}= \restric{\lder{\tilde W}{\Phi^*_A}}{q},
\label{relations der w}
\ee
--where now the ``$q$''-restriction means on the surface
$\{\Psi=\bar\Psi_q(\Phi,\Phi^*), \Psi^*=0\}$--
and of the explicit form \bref{aux quantum action} of $\tilde{W}$ in terms
of the original $W$, which indicates that dependence of
$\restric{\lder{\tilde W}{\Phi^*_A}}{q}$ on the fields
$\Phi$ always appears through the combination $(\Phi+\bar\Psi_q)$, yields
$$
  \Delta W_\Lambda=\left[
  W^A_{\,B}(\Phi+\bar\Psi_q,\Phi^*\gh\veps^2)\,
  \frac{\partial_r(\Phi+\bar\Psi_q)^B}{\partial\Phi^C}\,
  (\gh\veps^2)^C_{\,A} \right],
$$
with $W^A_{\,B}(\Phi,\Phi^*)$ given in terms of the original quantum
action $W$ by
\be
   W^A_{\,B}(\Phi,\Phi^*)=
 \frac{\partial_r\partial_l W}{\partial\Phi^B\partial\Phi^*_A}.
\label{wab}
\ee

On the other hand, differentiation of the quantum equations of
motion for $\Psi$ \bref{q equations motion}
$$
  \lder{\bar\Psi^A_q}{\Phi^B}=
  \lder{(\Phi+\bar\Psi_q)^C}{\Phi^B}\left(
  \frac{\partial_l\partial_r \gh Y}{\partial\Phi^C\partial\Phi^D}
  (\Phi+\bar\Psi_q,\Phi^*\gh\veps^2)\right)
  \gh O^{DA},
$$
allows to solve for the derivative of $\bar\Psi_q$ with respect $\Phi$ and
write
\be
  \Delta W_\Lambda(\Phi,\Phi^*)=
  \left[ W^A_{\,B}\,{\gh K}^{BC} (\gh O^{-1})_{CD}\,
  (\gh\veps^2)^D_{\,A} \right] (\Phi+\bar\Psi_q,\Phi^*\gh\veps^2),
\label{delta reg}
\ee
where the (inverse of the) operator ${\gh K}^{AB}$ is expressed as
$$
  ({\gh K}^{-1})_{AB}(\Phi,\Phi^*)=
  \left[(\gh O^{-1})_{AB}-\gh Y_{AB}(\Phi,\Phi^*)\right],
  \quad\quad\mbox{\rm with}\quad\quad
  \gh Y_{AB}=
  \frac{\partial_l\partial_r\gh Y}{\partial\Phi^A\partial\Phi^B}.
$$
The operator $({\gh K}^{-1})_{AB}$
appears in this way as the natural quantum,
antifield extension of the original operator
$(K^{-1})_{AB}$ \bref{k operator}. Further rewriting of the operator
${\gh K}^{AC}(\gh O^{-1})_{CB}$ in \bref{delta reg} as
\be
  {\gh K}^{AC}(\gh O^{-1})_{CB}=
  \left[\gh O^{AC}({\gh K}^{-1})_{CB}\right]^{-1}=
  \left(\delta^A_B-
  \gh O^{AC}\gh Y_{CB}\right)^{-1}\equiv(\delta_q)^A_{\,B}
\label{quantum delta}
\ee
finally yields
\be
  \Delta W_\Lambda(\Phi,\Phi^*)=
  \left[ W^A_{\,B}\, (\delta_q)^B_{\,C}\,(\gh\veps^2)^C_{\,A} \right]
  (\Phi+\bar\Psi_q,\Phi^*\gh\veps^2)
  \equiv\Omega(\Phi+\bar\Psi_q,\Phi^*\gh\veps^2).
\label{final delta reg}
\ee

Comparing then with the original (formal) computation of $\Delta W$
\be
   \Delta W= W^A_{\,B} \delta^B_{\,A},
\label{original delta w}
\ee
it is seen that nonlocal regularization acts
by essentially distorting the identity $\delta^A_{\,B}$
in \bref{original delta w} to a regulated expression
$(\delta_q)^A_{\,C}(\gh\veps^2)^C_{\,B}$
and by changing afterwards the arguments $(\Phi,\Phi^*)$ of the resulting
quantity $\Omega(\Phi,\Phi^*)$ to $(\Phi+\bar\Psi_q,\Phi^*\gh\veps^2)$.
In this sense, it proceeds in quite the same fashion
as usual regularization methods (Fujikawa regularization, PV,...),
i.e., by substituting in an appropriate way $\delta^A_{\,B}$ by some
other suitable expression.

On the other hand, the second term in the regulated quantum master
equation \bref{reg qme}, namely $(W_\Lambda,W_\Lambda)$, can be written,
using relations \bref{relations der w}, and in analogy with equation
\bref{s lambda 0} for $S_\Lambda$, as
$$
  \frac12(W_{\Lambda}, W_{\Lambda})=
  \restric{\rder{\tilde W}{\Phi^A}\lder{\tilde W}{\Phi^*_A}}{q}=
  \restric{\left(\rder{\tilde W}{\Phi^A}\lder{\tilde W}{\Phi^*_A}+
  \rder{\tilde W}{\Psi^A}\lder{\tilde W}{\Psi^*_A}\right)}{q}=
  \frac12\restric{(\tilde W, \tilde W)}{q},
$$
whereas use of the explicit form of $\tilde W$ \bref{aux quantum action}
and the canonical character of the transformation \bref{canonical transf}
yields
$$
  \restric{(\tilde W, \tilde W)}{q}=
  \restric{(W, W)(\Theta,\Theta^*)}{q}=
  (W,W)(\Phi+\bar\Psi_q,\Phi^*\gh\veps^2).
$$

Taking thus into account the final form of $\Delta W_\Lambda$
\bref{final delta reg} in terms of the quantity $\Omega$ and the above
result, expression \bref{reg qme} adopts the form
\be
   \gh A_\Lambda=
   \left[\Omega+\frac{i}{2\hbar}(W,W)\right]
  (\Phi+\bar\Psi_q,\Phi^*\gh\veps^2).
\label{final reg qme}
\ee
The regularized quantum master equation consists thus in a quantity, the
one in square brackets, entirely computed in terms of the original theory,
whereas all the dependence on the quantum on-shell auxiliary fields
$\bar\Psi_q$ appears only in the argument. This fact simplifies
considerably the study of this equation, so that in the end the knowledge
of the precise form of these quantum on-shell shadow fields results to be
unnecesary.

\subsection{Anomalies}

\hspace{\parindent}%
In this section, we will be mainly interested in the characterization of
anomalies in this framework. The anomaly question in this context has
already been considered in \cite{hand92}
for the specific examples of the chiral Schwinger model and QED.
In what follows, this topic is adressed in a
general way by investigating the form of the regularized quantum master
equation \bref{final reg qme}.

In the standard, locally regularized approaches, genuine anomalies are
interpreted as obstructions to the local solvability of the (complete)
quantum master equation. Nonlocal regularization, however, due to its
nonlocal character, must be supplemented with an interpretation of what
is meant by ``local solvability''. A general criterion comes from
the fact that nonlocality of the approach is only part of the
regularization method, so that for a local, renormalizable theory,
in the limit $\Lambda^2\rightarrow\infty$, all the divergent and finite
quantities arising upon computation of $\Delta W_\Lambda$ in
\bref{final delta reg} should become local.
Therefore, acceptable solutions for $W_\Lambda$, or equivalently, for the
counterterms $M_{p,\Lambda}$, in this framework must be entirely
constructed, before taking the limit $\Lambda^2\rightarrow\infty$, from the
smearing operator $\gh\veps^A_{\,B}$ \bref{smearing op}
in such a way that any sort of nonlocality disappear in the
unregularized limit. If such a choice does not exist, some of the
original, local symmetries may become anomalous.

The regulated quantum master equation \bref{final reg qme} can in fact be
naturally decomposed into its divergent part and its finite part
when $\Lambda^2\rightarrow\infty$, the latter being indicated as
$\left[\gh A_\Lambda\right]_0$.
``Local'' solvability of the divergent part, as already stressed in the
original references \ct{emkw91,kw92,kw93}, defines an interesting problem
on its own, which is not going to be adressed in the present study. It is
worth to note, however, that the validity of this regularization method,
in view of the renormalization of the theory, heavily relies on the
fulfillment of this condition.

Anomaly issues, instead, are encoded in the finite part,
$\left[\gh A_\Lambda\right]_0$, of \bref{final reg qme}.
More concretely, the expression of the anomaly predicted by the regularized
BRST Ward identity \bref{reg ward ident} should be considered, as in other
regularization approaches, the value of this finite part in the limit
$\Lambda^2\rightarrow\infty$
$$
   \gh A\equiv \lim_{\Lambda^2\rightarrow\infty}
   \left[\Omega+\frac{i}{2\hbar}(W,W)\right]_0
  (\Phi+\bar\Psi_q,\Phi^*\gh\veps^2),
$$
which, after taking into account in the arguments the limits
$\bar\Psi_q\rightarrow 0$, $\gh\veps^2\rightarrow 1$ when
$\Lambda^2\rightarrow\infty$, becomes
\be
   \gh A=\left[(\Delta W)_R +\frac{i}{2\hbar}(W,W)\right](\Phi,\Phi^*),
\label{fin anomaly 2}
\ee
with the regularized value of $\Delta W$ defined as
\be
   (\Delta W)_R\equiv
   \lim_{\Lambda^2\rightarrow\infty} [\Omega]_0,
\label{delta w r omega}
\ee
and where, from now on, $W$ will stand for the finite part of the actual
quantum action \bref{quantum action}, i.e., the one without
the divergent parts of the counterterms needed for renormalization.
In this way, as anticipated, due to the precise dependence of the regulated
quantum master equation in the quantum on-shell shadow fields,
computation of the anomaly can be performed without the precise
knowledge of their form.

For practical perturbative calculations, it is convenient to
analyze the $\hbar$ expansion of \bref{fin anomaly 2} in order to
recognize the expressions of the $p$-loop obstructions appearing in
the regularized BRST Ward identity \bref{reg ward ident}. Consider then
definition \bref{final delta reg} for $\Omega$. The operator
$(\delta_q)^A_{\,B}$ \bref{quantum delta} involved in its computation can
first be written, using \bref{quantum action 2}, as
$$
  (\delta_q)^A_{\,B}=
  \left[(\delta^{-1}_\Lambda)^A_{\,B}
  -\sum^\infty_{p=1}{\hbar}^p (\gh O M_p)^A_{\,B}\right]^{-1},
$$
where $(\delta_\Lambda)^A_{\,B}$ is defined by
\be
  (\delta_\Lambda)^A_{\,B}=
  \left(\delta^A_B-\gh O^{AC}\gh I_{CB}\right)^{-1}=
   \delta^A_{\,B}+\sum_{n=1}
  \left(\gh O^{AC}\gh I_{CB}\right)^{n},
  \quad\mbox{with}\quad
   \gh I_{AB}=\frac{\partial_l\partial_r \gh I}
    {\partial\Phi^A\partial\Phi^B},
\label{delta lambda}
\ee
and where $(\gh O M_p)^A_{\,B}$ stands for the shorthand notation
$$
  (\gh O M_p)^A_{\,B}= \gh O^{AC} (M_p)_{CB},\quad\mbox{with}\quad
   (M_p)_{AB}=\frac{\partial_l\partial_r M_p}
    {\partial\Phi^A\partial\Phi^B}.
$$
Further expansion in powers of $\hbar$ results finally in
\be
  (\delta_q)^A_{\,B}= (\delta_\Lambda)^A_{\,B}+\sum_{p=1}\hbar^p
  \sum_{\mbox{\small$\begin{array}{cc}
  0<p_1,\ldots,p_j\leq p\\
  p_1+\ldots+p_j=p
  \end{array}$}}
  \left( \delta_\Lambda(\gh O M_{p_1}) \delta_\Lambda\ldots
  \delta_\Lambda(\gh O M_{p_j}) \delta_\Lambda\right)^A_{\,B},
\label{final delta q}
\ee
where the sum runs over all permutations of the indices $p_1,\ldots,p_j$
such that their sum is $p$, the corresponding power of $\hbar$.

Plugging now the $\hbar$ expansion of $W^A_{\,B}$ \bref{wab},
inferred from expansion \bref{quantum action} for $W$,
\be
  W^A_{\,B}= S^A_{\,B}+
  \sum^\infty_{p=1}{\hbar}^p (M_p)^A_{\,B},
\label{exp wab}
\ee
and expansion \bref{final delta q} in expression \bref{final delta reg} for
$\Delta W_\Lambda$ allows to recognize the different
coefficients in the $\hbar$ expansion of $\Omega$,
$\Omega=\sum_0\hbar^p \Omega_p$. The lowest order term, $\Omega_0$,
appears to be
\be
   \Omega_0=
  \left[S^A_{\,B}(\delta_\Lambda)^B_{\,C}(\gh\veps^2)^C_{\,A}\right],
\label{delta s}
\ee
while for $\Omega_p$, it is
\bea
   &&\Omega_p=
  \left[(M_p)^A_{\,B}(\delta_\Lambda)^B_{\,C}(\gh\veps^2)^C_{\,A}\right]
\nonumber\\
   &&+ \sum_{\mbox{\small$\begin{array}{cc}
   0<p_1,\ldots,p_j\leq p\\p_1+\ldots+p_j=p
   \end{array}$}}
   \left[
   S^A_{\,B} \left( \delta_\Lambda(\gh O M_{p_1}) \delta_\Lambda\ldots
   \delta_\Lambda(\gh O M_{p_j}) \delta_\Lambda\right)^B_{\,C}
  (\gh\veps^2)^C_{\,A}\right]
\nonumber\\
   &&+ \sum_{\mbox{\small$\begin{array}{cc}
   0<p_1,\ldots,p_j< p\\p_1+\ldots+p_j=p
   \end{array}$}}\left[
   (M_{p_1})^A_{\,B} \left( \delta_\Lambda(\gh O M_{p_2})
   \delta_\Lambda\ldots \delta_\Lambda(\gh O M_{p_j})
   \delta_\Lambda\right)^B_{\,C} (\gh\veps^2)^C_{\,A}\right].
\label{delta mp}
\eea

By finally dropping out the divergent part of these expressions as part of
the renormalization procedure and taking the limit
$\Lambda^2\rightarrow\infty$ in the remaining finite expressions,
as indicated by \bref{delta w r omega}, what should be taken as
the regularized values of $\Delta S$ and $\Delta M_p$,
$(\Delta S)_R$ and $(\Delta M_p)_R$, is obtained
\be
   (\Delta S)_R\equiv
   \lim_{\Lambda^2\rightarrow\infty} [\Omega_0]_0,\quad\quad
   (\Delta M_p)_R\equiv
   \lim_{\Lambda^2\rightarrow\infty} [\Omega_p]_0,\quad p\geq1.
\label{reg values}
\ee
These are precisely the values appearing in the regularized analogs of
expressions \bref{first order} and \bref{higher order},
obtained from \bref{fin anomaly 2} upon expanding in $\hbar$
\bea
     \gh A_1&=&(\Delta S)_R+i(M_1, S),
\label{reg first order}\\
     \gh A_p&=&(\Delta M_{p-1})_R
     +\frac{i}2\sum^{p-1}_{q=1} (M_q,M_{p-q}) +i(M_p, S),
     \quad\quad p\geq 2,
\label{reg higher order}
\eea
which should be considered the form of the $p$-loop anomaly
provided by the nonlocally regularized BRST Ward identity
\bref{reg ward ident}.

Some comments are finally in order about the above expressions.
First of all, it is not difficult to see that $(\Delta S)_R$
and, as a consequence, the one-loop anomaly
\bref{reg first order}, satisfy the usual Wess-Zumino consistency
condition \cite{wz}. Indeed, consider the quantity $\Delta S_\Lambda$,
which can be written in terms of $\Omega_0$ \bref{delta s} as
$$
  \Delta S_\Lambda(\Phi,\Phi^*)=\Omega_0
  (\Phi+\bar\Psi_0,\Phi^*\gh\veps^2),
$$
with $\bar\Psi_0$ the classical on-shell shadow fields. The algebraic
definitions of the antibracket \bref{antibracket} and of the operator
$\Delta$ \bref{delta op}, together with the fact that $S_\Lambda$ verifies
the classical master equation \bref{s lambda 0}, allow then to conclude
the condition
$$
  (\Delta S_\Lambda, S_\Lambda)(\Phi,\Phi^*)=0 \Rightarrow
  (\Omega_0, S) (\Phi+\bar\Psi_0,\Phi^*\gh\veps^2) =0,
$$
where in writing the right hand side, relations
\bref{relations der} and
$\restric{\lder{\tilde S}{\Psi^*_A}}{}=(\bar\Psi_0, S_\Lambda)$
have been used. This condition holds as well for the finite part of
$\Omega_0$, $[\Omega_0]_0$, so that in the limit
$\Lambda^2\rightarrow\infty$ the consistency condition arises
$$
   \lim_{\Lambda^2\rightarrow\infty}([\Omega_0]_0, S)
   (\Phi+\bar\Psi_0,\Phi^*\gh\veps^2)=
   ((\Delta S)_R,S)(\Phi,\Phi^*)=0.
$$
Nonlocal regularization is thus a consistent regularization procedure, in
the sense that it yields one-loop anomalies which verify the Wess-Zumino
consistency condition.

On the other hand, expression \bref{reg higher order} for the higher order
loop terms in the quantum master equation can naturally be interpreted, as
already suggested in the introduction, as the one-loop
corrections generated by the counterterms $M_k$, $k< p$, to the $p$-loop
anomaly. The fact that $(\Delta M_{p})_R$ \bref{delta mp}
appears as a functional trace over a certain operator, one of the most
typical characteristics of one-loop corrections, further supports this
conclusion. The obtained results indicate then that
the BRST Ward identity \bref{ward identity} and its regulated version
\bref{reg ward ident} proposed in the FA framework are incomplete and
that, at two and higher loops, fundamental pieces are missed in the
resulting quantum master equation. This incompleteness, however, should
only be considered as a deficiency in the derivation of the right hand
side of the BRST Ward identity \bref{ward identity} and not as an
inability of nonlocal regularization to deal with two and higher loop
corrections. These conclusions are strongly supported by the example
presented below, for which computation of the universal two-loop anomaly
through the BRST variation of the nonlocally regulated effective action
$\Gamma$ leads to the right answer, while use of the two-loop quantum
master equation reproduces only the contribution to the universal two-loop
anomaly of a standard one-loop counterterm absorbing part of the one-loop
anomaly.

\section{An example: chiral $W_3$ Gravity}

\hspace{\parindent}
In this section, the use of the nonlocally regulated FA formalism is
exemplified by calculating the one and two
loop anomalies for chiral $W_3$ gravity in this framework. The forms
obtained for these quantities are in complete agreement with previous
results obtained in the literature
\cite{mat89,hull91,ssn91,prs91,hull93}, thereby showing the
nonlocally regulated FA formalism as a suitable candidate for further
study and characterization of the structure of higher loop BRST anomalies.
We closely follow the conventions and notations of ref.\,\cite{vp94}, to
which the reader is referred for the explicit construction of the proper
solution of the master equation in the classical and gauge-fixed basis.

\subsection{Nonlocalization of the proper solution}

\hspace{\parindent}
Chiral $W_3$ gravity \cite{hull90a} consists in a system of $D$ scalar
fields $\phi^i$, $i=1,\ldots, D$, coupled to
gauge fields $h$ and $B$ through the spin-2 and spin-3 currents
\be
  T=\frac12(\partial\phi^i)(\partial\phi^i),\quad\quad
  W=\frac13 d_{ijk} (\partial\phi^i) (\partial\phi^j) (\partial\phi^k),
\label{currents}
\ee
where we are using the notations
$$
  \partial=\partial_+,\quad\quad \bar\partial=\partial_-,\quad\quad
  x^{\pm}=\frac1{\sqrt2}(x^1\pm x^0),
$$
and where $d_{ijk}$ is a constant, totally symmetric tensor
satisfying the identity
$$
  d_{i(jk}d_{l)mi}=k\delta_{(jl}\delta_{k)m},
$$
in terms of an arbitrary, but fixed parameter $k$.
The classical action obtained in this way
$$
   S_0=\int\dif^2 x\,
   \left[-\frac12(\partial\phi^i)(\bar\partial\phi^i)+h\,T+B\,W \right],
$$
with $\dif^2 x=\dif x^0\dif x^1=\dif x^+\dif x^-$, is invariant under
well-known spin-2 and spin-3 gauge symmetries having an open algebra.

In the gauge-fixed basis of fields and antifields,
the proper solution of the classical master equation for this system,
as constructed in \cite{vp94}, is given by%
\footnote{The free parameter $\alpha$ considered in \cite{vp94} is taken
here equal to $0$ for simplicity.}
\bea
   S=\int\dif^2 x\hspace{-5mm}&&\left\{
   \left[-\frac12(\partial\phi^i)(\bar\partial\phi^i)
   +b(\bar\partial c)+ v(\bar\partial u)\right]\right.
\nonumber\\
   &&+\phi^*_i\left[c(\partial\phi^i)+
   u d_{ijk} (\partial\phi^j) (\partial\phi^k)
   -2 k b (\partial u) u (\partial\phi^i)\right]
\nonumber\\
   &&+b^*\left[-T +2 b(\partial c)+ (\partial b) c
   + 3 v(\partial u) + 2(\partial v) u \right]
\nonumber\\
   && +v^*\left[-W +2 k T b(\partial u) +2 k \partial(T b u)
   + 3 v(\partial c) + (\partial v) c \right]
\nonumber\\
   && \left.
   +c^*\left[ (\partial c) c
   +2 k T (\partial u) u\right]
   +u^*\left[2(\partial c) u- c(\partial u)\right]\right\}
\nonumber\\
   &=& \gh S(\Phi) +\Phi^*_A R^A(\Phi),
\label{w3gfa}
\eea
where $\{c, u\}$ are the ghosts corresponding to spin-2 and spin-3 gauge
symmetries; $\{b,v\}$, their associated antighosts; and
$\{\phi^*_i, c^*, u^* ,b^*, v^*\}$, the corresponding antifields. This
action is obtained by first constructing the proper solution in the
classical basis of fields and antifields
$\{\phi^i, h, B, c, u;\phi^*_i, h^*, B^*, c^*, u^*\}$ and
performing afterwards the canonical transformation to the gauge-fixed basis
$$
   \{h, h^*, B, B^*\}\rightarrow
   \{b=h^*, b^*=-h, v=B^*, v^*=-B\}.
$$
It is worth to note that the resulting gauge-fixed action contains
no interaction, i.e., $\gh S(\Phi)=F(\Phi)$ and $I(\Phi)=0$, so that
the requirement for applying nonlocal regularization to this model is
satisfied. Interactions can be considered to be contained in the
antifield dependent part, $\Phi^*_A R^A(\Phi)$, antifields acting then
as a sort of coupling constants on which expansions can be performed.

The first step towards nonlocalization of the proper solution
\bref{w3gfa} is the identification of the kinetic operator
$\gh F_{AB}$ in \bref{original action} for the propagating fields
$\Phi^A=\{\phi^i; b, v; c, u\}$. In this basis, in which from now on
all matrices are going to be expressed, its explicit expression reads
$$
 \gh F_{AB}=
 \left(\begin{array}{ccc}
    \partial\bar\partial\,\delta_{ij} & 0 & 0 \\
    0 & 0 & \unity\,\bar\partial \\
    0 &  \unity\,\bar\partial  &  0
  \end{array}\right),
$$
where $\unity$ stands for the identity in the spin 2 (spin 3) ghost
sector. Introducing then an operator $(T^{-1})^{AB}$ of the form
$$
 (T^{-1})^{AB}=
 \left(\begin{array}{ccc}
    \delta^{ij} & 0 & 0 \\
    0 & 0 & \unity\,\partial\\
    0 & \unity\,\partial  & 0
  \end{array}\right),
$$
a suitable regulator, quadratic in space-time derivatives, arises
\be
   \gh R^A_{\,B}=(T^{-1})^{AC}\gh F_{CB}=
   \partial\bar\partial\,\delta^A_{\,B},
\label{w regulator}
\ee
with $\delta^A_{\,B}$ the identity in the complete space of fields.
The corresponding smearing and shadow kinetic operator are afterwards
constructed from \bref{w regulator} using the general expressions
\bref{smearing op} and \bref{shadow kinetic op}, resulting in
\be
   \gh\veps^A_{\,B}=
   \exp\left(\frac{\partial\bar\partial}{2\Lambda^2}\right)
   \delta^A_{\,B}\equiv\gh\veps \,\delta^A_{\,B}, \quad\quad
   \gh O^{-1}_{AB}= (\gh\veps^2-1)^{-1}\gh F_{AB},
\label{w smearing op}
\ee
whereas $\gh O^{AB}$ takes the form
\be
   \gh O^{AB}=
   \left(\begin{array}{ccc}
   \gh O & 0 & 0 \\
   0 & 0 & \unity\gh O \partial\\
   0 & \unity \gh O \partial  & 0
   \end{array}\right),
   \quad\quad\mbox{with} \quad\quad
   \gh O\equiv \frac{(\gh\veps^2-1)}{\partial\bar\partial}=
   \int^1_0\frac{\dif t}{\Lambda^2}\,
   \exp\left(t\frac{\partial\bar\partial}{\Lambda^2}\right).
\label{w shadow prop}
\ee

Nonlocalization of the proper solution \bref{w3gfa} (or of a suitable
quantum extension $W$ of it, if counterterms are needed) would now proceed
as described in section 2, that is, by introducing the shadow fields and
antifields $\{\Psi^A,\Psi^*_A\}$, constructing from them and the above
objects the auxiliary proper solution \bref{auxiliary proper} (or its quantum
extension \bref{aux quantum action}), and
substituting the shadow fields by the solutions of their equations of
motion, while putting their antifields to zero. In the present case,
due to the absence of the classical interaction term $I(\Phi)$ and the
highly nonlinear form of the BRST transformations $R^A(\Phi)$ in
\bref{w3gfa}, it is clear that the classical shadow fields should be
solved perturbatively in antifields. However, as analyzed in the previous
section, computation of anomalies from the regularized form of the quantum
master equation can be completely performed without this information,
whereas the two-loop anomaly calculation, relying on diagrammatics, only
needs, as indicated in section 2, the form of the auxiliary proper solution
\bref{auxiliary proper}. For this reason, we skip this calculation
and leave it as an exercise for the interested reader.

\subsection{One-loop anomaly}

\hspace{\parindent}
Let us now calculate the one-loop anomaly according to the prescription
\bref{reg first order} presented in the previous section.
The main ingredients to compute $(\Delta S)_R$ are, as indicated by
\bref{reg values}, \bref{delta s}, \bref{delta lambda}, the objects
$S^A_{\,B}$, $\gh O^{AB}$, $\gh I_{AB}$ and $(\gh\veps^2)^A_{\,B}$.

The operators $(\gh\veps^2)^A_{\,B}$ and $\gh O^{AB}$ have been previously
constructed and are given
by \bref{w smearing op} and \bref{w shadow prop}, respectively.
On the other hand, expressions for
$S^A_{\,B}$, \bref{exp wab}, \bref{wab}, and
$\gh I_{AB}$, \bref{delta lambda},
are constructed entirely in terms of the original proper solution $S$.
In the present example, we have for $S^A_{\,B}$
$$
  S^A_{\,B}=
  \frac{\partial_r\partial_l S}{\partial\Phi^B\partial\Phi^*_A}=
$$
\be
   \left(\begin{array}{ccccc}
    c^i_j \partial & -2k(\partial u) u (\partial\phi^i) & 0 &
    (\partial\phi^i) & u^i\\
    -(\partial\phi_j)\partial & -(c\partial)_2 & -2(u\partial)_{3/2} &
    (b\partial)_1 & 3(v\partial)_{1/3}  \\
    -u_j\partial & -2k[ T (u\partial)_2+u(\partial T)] & -(c\partial)_3 &
    3(v\partial)_{1/3} &  4 k (bT\partial)_{1/2} \\
    2k(\partial u) u (\partial\phi_j) \partial &  0 & 0 &
    -(c\partial)_{-1} & -2 k T (u\partial)_{-1}\\
    0 & 0 & 0 & -2 (u\partial)_{-1/2} & -(c\partial)_{-2}
   \end{array}\right),
\label{wsab}
\ee
with $T$ the spin-2 current in \bref{currents}; $c^i_j$ and $u^i$,
the operators
\bea
     c^i_j&=& \left[c\delta^i_j -2k b(\partial u) u\delta^i_j+
     2 u d^i_{\,jk}(\partial\phi^k)\right],
\nonumber\\
     u^i&=& d^i_{\,jk} (\partial\phi^j) (\partial\phi^k)
     -2 k \left[b (\partial u) (\partial\phi^i)
                +(b (\partial\phi^i) u\partial)_1\right],
\nonumber
\eea
and where $(F(\Phi,\Phi^*)\,\partial)_n$ stands for the shorthand notation
\be
   (F\,\partial)_n= F\,\partial+ n (\partial F),
   \quad\quad (F\,\partial)_n^\dagger=
   -[F\,\partial+ (1-n) (\partial F)]=-(F\,\partial)_{1-n}.
\label{f partial}
\ee

In much the same way, the operator $\gh I_{AB}$ reads in this case
$$
  \gh I_{AB} =
  \frac{\partial_l\partial_r}{\partial\Phi^A\partial\Phi^B}
  \left[\Phi^*_C R^C(\Phi)\right]=
$$
\be
  \left(\begin{array}{ccccc}
  \partial\,h^*_{ij}\,\partial & (g^*_i\partial)_1  & 0 &
  -(\phi^*_i\partial)_1 & (q^*_i\partial)_1\\
  g^*_j\,\partial & 0 & 0 & (b^*\partial)_{-1} & r^*\\
  0 & 0 & 0 & 2(v^*\partial)_{-1/2} & (b^*\partial)_{-2}\\
  -\phi^*_j\,\partial & (b^*\partial)_2 & 2 (v^*\partial)_{3/2} &
  2(c^*\partial)_{1/2} & -3(u^*\partial)_{2/3}\\
  q^*_j \, \partial & - (r^*)^\dagger & (b^*\partial)_{3} &
  -3(u^*\partial)_{1/3} & 2(p^*\partial)_{1/2}
   \end{array}\right),
\label{w tilde iab}
\ee
where the linear quantities in the antifields
$ h^*_{ij}$, $b^*_i$, $q^*_i$, $r^*$ and $p^*$
are given by
\bea
   h^*_{ij}&=&\left\{ \delta_{ij}
   \left[ b^* + 2kb(u(\partial v^*)-v^*(\partial u))
   +2k c^*u(\partial u)\right]
   -2 d_{ij}^{\,\,\,\,k}
   \phi^*_k u +2 v^* d_{ijk} (\partial\phi^k)\right\},
\nonumber\\
    g^*_i&=& 2k\left[\phi^*_i(\partial u) u+
   (v^*(\partial u) -u(\partial v^*))(\partial\phi_i)\right],
\nonumber\\
    q^*_i&=&\left\{ -2 \phi^*_j d^j_{\,ik}(\partial\phi^k)+
    2k\left[\left( \partial(v^*b (\partial\phi_i)+u\phi^*_i b)\right)+
    b(\partial\phi_j)(v^*\partial)_1+\phi^*_i b (u\partial)_1\right]
    \right\},
\nonumber\\
    r^*&=&
    2k\left[T(v^*\partial)_{-1}
    -\phi^*_i(\partial\phi^i)(u\partial)_{-1}\right],
\nonumber\\
    p^*&=& 2k\left[T c^*-\phi^*_i(\partial\phi^i)b\right].
\nonumber
\eea

Linearity of the proper solution \bref{w3gfa} in antifields
leads thus to operators
$S^A_{\,B}$ \bref{wsab} and $\gh I_{AB}$ \bref{w tilde iab}
independent of and linear in the antifields, respectively, so that
expression \bref{delta s} for $\Omega_0$ becomes
an usual antifield expansion, after plugging in it
expansion \bref{delta lambda} for $(\delta_\Lambda)^A_{\,B}$ adapted to
this case. Dimensional analysis in terms of the
combination $d-j$ --``engineering'' dimension minus spin--
appears then to be very useful to figure out the relevant terms in
this antifield expansion for $\Omega_0$. Indeed, the value of the $d-j$
combination for the relevant quantities involved in its computation
\be
  (d-j)[\Phi^A,\,\partial]=0, \quad\quad
  (d-j)[\Phi^*_A,\,\Lambda^2]=2,
\label{d minus j}
\ee
and the fact that $(\Delta S)_R\sim\Omega_0$ is the integral of a
quantity of dimension $d=2$ and spin $j=0$, or $d-j=2$, constraint the
possible terms arising in its calculation to be necessarily of the form
$$
   \Lambda^{-2n} (\Phi^*)^{2(n+1)} F_n(\Phi;\partial),
   \quad\quad n=-1, 0, 1, \ldots,
$$
so that only the terms $n=-1, 0$ --the divergent,
antifield independent piece and the finite term, linear in antifields--
are really relevant. Collecting then all these facts together, it is
concluded that the relevant terms
in expansion \bref{delta s}, \bref{delta lambda} for $\Omega_0$ are
\be
  \Omega_0= \left[ \gh\veps^2\, S^A_{\,A}\right]+
  \left[ \gh\veps^2\, S^A_{\,B}\, \gh O^{BC}\,\gh I_{CA}\right]+
  O\left(\frac{(\Phi^*)^2}{\Lambda^2}\right).
\label{antifield exp delta}
\ee

The antifield independent term in expansion \bref{antifield exp delta},
encoding potential divergencies, is just the functional trace in the
continuous indices of $S^A_{\,A}$ weighted with the ``damping'' operator
$\gh\veps^2$, resulting in
$$
  \left[ \gh\veps^2 \,S^A_{\,A}\right]=
  {\rm Tr}\left[\gh\veps^2
  \left(c^i_i\partial -(c\partial)_2 -(c\partial)_3 -(c\partial)_{-1}
  -(c\partial)_{-2}\right) \right]=
  {\rm Tr}\left[\gh\veps^2
  \left(c^i_i\partial -4(c\partial)_{1/2}\right)\right]=0,
$$
since $\gh T(F,n)={\rm Tr}\left[\gh\veps^2(F\partial)_{n}\right]=0$, as
stated by computation \bref{integral 0} in the appendix.
The vanishing of this term indicates that, upon expanding in $\Lambda^2$,
finite contributions to $[\Omega_0]_0$ will only come from the second term
in \bref{antifield exp delta}.

The next step is then the determination of the diagonal elements of the
matrix $S^A_{\,B} \gh O^{BC}\gh I_{CD}$. A straightforward calculation
yields $$
  {\rm diag}(S^A_{\,B} \gh O^{BC}\tilde I_{CD})=
  (A^i_j, \,A^b_b\, , \,A^v_v\, , \,A^c_c\, , \,A^u_u),
$$
with the above operators given by
\bea
    A^i_j&=&c^{ik}\,\partial\,\gh O\,\partial \,h^*_{kj}\,\partial
         +2k(\partial u)u(\partial\phi^i)\, \gh O\,\partial\,
         \phi^*_j\,\partial
         +(\partial\phi^i) \,\gh O\,\partial\, g^*_j\, \partial,
\label{aij}\\
    A^b_b&=& -(c\partial)_2 \,\gh O\,\partial\,(b^*\partial)_2
             -(\partial\phi^i)\, \partial\,\gh O\,\partial\,g^*_i\,\partial
            +2(u\partial)_{3/2}\, \gh O\,\partial\, (r^*)^\dagger,
\label{abb}\\
    A^v_v&=& -(c\partial)_3 \,\gh O\, \partial\, (b^*\partial)_3
             -4k\left[T(u\partial)_2+u(\partial T)\right]\,
             \gh O\,\partial\,(v^*\partial)_{3/2},
\label{avv}\\
    A^c_c&=& -(c\partial)_{-1} \,\gh O\, \partial\, (b^*\partial)_{-1}
         -2k(\partial u)u(\partial\phi^i)\, \partial \,\gh O\,
         (\phi^*_i\partial)_1
          -4k\,T(u\partial)_{-1}\,\gh O\,\partial\,(v^*\partial)_{-1/2},
\label{acc}\\
    A^u_u&=& -(c\partial)_{-2}\, \gh O\,\partial\,(b^*\partial)_{-2}
              -2(u\partial)_{-1/2}\, \gh O\,\partial\, r^*.
\label{auu}
\eea
The expression of $(\Delta S)_R$ will then be, according to
\bref{reg values}
\be
   (\Delta S)_R= \lim_{\Lambda^2\rightarrow\infty} [\Omega_0]_0=
   \lim_{\Lambda^2\rightarrow\infty}
  {\rm Tr} \left[\gh\veps^2
  \left(A^i_i+ A^b_b+ A^v_v+ A^c_c+ A^u_u\right)\right].
\label{1 anomaly 1}
\ee

At this point, it is interesting to note that all the above traces appear,
or can be written, as particular cases of the general expression
\be
  \gh T(F,G; n,m)=
  \lim_{\Lambda^2\rightarrow\infty}
  {\rm Tr} \left[\gh\veps^2\,
  F\,\partial^n \,\gh O\,\partial\, G\,\partial^m\right],
\label{general trace}
\ee
whose explicit computation, performed in the appendix, yields
\be
  \gh T(F,G; n,m)= \frac{-i}{2\pi}
   \left[\sum_{k=0}^m {m \choose k}
   \frac{(-1)^k}{n+m+1-k}\left(1-\frac1{2^{n+m+1-k}}\right)\right]
   \int\dif^2 x\, F\,\partial^{n+m+1}\, G.
\label{general trace bis}
\ee
This result can now be extensively used to calculate the functional trace
\bref{1 anomaly 1}. For example, the first term in $A^i_j$ \bref{aij},
corresponding to the matter fields, reads in the above notation
$\gh T(c^{ij},h^*_{ji}; 1,1)$,
giving in this way the well-known contribution
\be
   (\Delta S)_R^{(i)}=\frac{i}{24\pi} \int\dif^2 x\,
   c^{ij}\,\partial^3\, h^*_{ij},
\label{contribution 1}
\ee
which besides the usual $W_2$ one-loop anomaly
$\frac{iD}{24\pi} \int\dif^2 x\, c\,\partial^3\, b^*$ contains some new
terms in the rest of antifields. The remaining contributions
which contain explicitly the fields $c$ and $b^*$, coming from the
ghosts field entries \bref{abb}--\bref{auu}, share the generic form
$$
  \lim_{\Lambda^2\rightarrow\infty}
 {\rm Tr} \left[\gh\veps^2\,
(c\partial)_j\, \gh O\,\partial\, (b^*\partial)_j\right]=
  j^2\gh T((\partial c),(\partial b^*);0 ,0)
   +\gh T(c,b^*;1 ,1)
$$
$$
  +j\left[\gh T((\partial c),b^*;0 ,1)+
  \gh T(c,(\partial b^*);1,0)\right]=
 \frac{i}{24\pi} (6j^2-6j+1) \int\dif^2 x\, c\,\partial^3\, b^*,
$$
where $j$ is the spin of the field associated with the entry, i.e.,
$j=(2, 3, -1, -2)$. All together, they add up to the well-known result
\be
 (\Delta S)_R^{(ii)}=
 \frac{-100\,i}{24\pi} \int\dif^2 x\, c\,\partial^3\, b^*.
\label{contribution 2}
\ee

The remaining contributions contain either $v^*$ or $\phi^*_i$. Proceeding
in much the same way by using the general result
\bref{general trace bis}, the $v^*$ terms are computed to be
\bea
 &\displaystyle{(\Delta S)_R^{(iii)}+ (\Delta S)_R^{(iv)}=
 \frac{ik}{2\pi} \int\dif^2 x\,
   (v^*(\partial u) -u(\partial v^*))(\partial\phi^i)(\partial^3\phi^i)}&
\nonumber\\
 &\displaystyle{+\frac{ik}{6\pi} \int\dif^2 x\,T\left[
    5 (\partial^3 u) v^*
    -12 (\partial^2 u)(\partial v^*)
    +12 (\partial u)(\partial^2 v^*)
    -5 u(\partial^3 v^*)\right]}&,
\label{contribution 3}
\eea
while the $\phi^*_i$ contribution results in
\be
   (\Delta S)_R^{(v)}=
    \frac{-ik}{6\pi} \int\dif^2 x\,\phi^*_i\left[
     6\partial\left(u (\partial u) (\partial^2\phi^i)\right)
     +9 (\partial^2 u) (\partial u) (\partial\phi^i)
     +8  u (\partial^3 u) (\partial\phi^i) \right].
\label{contribution 4}
\ee

In summary, expressions \bref{contribution 1}, \bref{contribution 2},
\bref{contribution 3} and \bref{contribution 4} together constitute
the form of $(\Delta S)_R$, or
of the complete consistent one-loop anomaly if no counterterm $M_1$ is
considered, provided by the nonlocally regularized FA formalism. Its form
is in complete agreement with the PV calculation performed in \cite{vp94},
for the case $\alpha=0$, and with earlier results in the literature
\cite{hull91,ssn91,prs91,hull93}, thus indicating that expression
\bref{reg first order} correctly reproduces one-loop anomalies.

\subsection{Counterterm corrections to the one and two-loop anomalies}

\hspace{\parindent}%
Contributions $(\Delta S)_R^{(i)}$ \bref{contribution 1} and
$(\Delta S)_R^{(ii)}$ \bref{contribution 2} to the complete
expression of $(\Delta S)_R$ are seen to contain a universal
(gravitational) one-loop anomaly, depending only on the spin 2
ghost $c$ and antifield $b^*$ (or gauge field $h$ when working in the
classical basis),
\be
  (\Delta S)_{R,{\rm univ}}=
 \frac{i(D-100)}{24\pi} \int\dif^2 x\, c\,\partial^3\, b^*,
\label{univ anom}
\ee
a mixed spin 2-spin 3 anomaly%
\footnote{It should be noted that the $k^2$--proportional terms
present in $(\Delta S)_R^{(i)}$ \bref{contribution 1} amount to a total
derivative. Therefore they are dropped out from \bref{mixed anomaly}.}
\be
  (\Delta S)_{R,{\rm mix}}= \frac{ikD}{12\pi}
  \int\dif^2 x\,\left\{
  c \,\partial^3\,
  \left[b(u(\partial v^*)-v^*(\partial u)) + c^*u(\partial u)\right]
  - [b(\partial u) u]\, \partial^3\, b^*\right\}
\label{mixed anomaly}
\ee
plus some extra terms containing either matter fields $\phi^i$ and/or
antifields $\phi^*_i$, which add to contributions
\bref{contribution 3}, \bref{contribution 4} in order to define the
complete matter dependent one-loop anomaly.

The appearence of the mixed spin 2-spin 3 anomaly \bref{mixed anomaly} can
be traced back \ct{hull90a} to the non-primary character of the total spin
3 current $W+W_{\rm gh}$ --appearing in \bref{w3gfa} as the $v^*$
coefficient-- with respect the total energy momentum tensor $T+T_{\rm gh}$
--the $b^*$ coefficient in \bref{w3gfa}.
Such part of the anomaly, however, can be absorbed by modifying in a
suitable way the definition of the total spin 3 current and of the BRST
transformation for the ghost $c$ \ct{prs91,hull93}, which in the present
notation amounts to add the finite one-loop counterterm
\bea
   M_1= \beta \int\dif^2 x\hspace{-5mm}&&\left\{
   v^*\left[ 2 u(\partial^3 b) +9 (\partial u)(\partial^2 b)
    + 15 (\partial^2 u)(\partial b) + 10 (\partial^3 u) b \right]\right.
\nonumber\\
   &&\left.+c^*\left[2 u(\partial^3 u)- 3(\partial u)(\partial^2 u)
   \right]\right\},
\label{m1 counterterm}
\eea
with $\beta=\frac{-kD}{192\pi}$, and define the quantum action
\bref{quantum action} as $W=S+\hbar M_1$.

The addition of such counterterm results in the following modifications
with respect the original anomalies given by $S$ \bref{w3gfa}. On the one
hand, the expression of the complete one-loop anomaly $\gh A_1$ changes
according to \bref{reg first order}. The net effect, apart from the
absence of the mixed spin 2-spin 3 anomaly \bref{mixed anomaly}, is a
modification of the contribution $(\Delta S)_R^{(iv)}$ in
\bref{contribution 3} to
\bea
   \widehat{(\Delta S)}_R^{(iv)}=
   \frac{ik}{6\pi} \int\dif^2 x\,T&&\hspace{-5mm}\left[
    \left(5+\frac{D}{16}\right) (\partial^3 u) v^*
    -\left(12+\frac{3D}{32}\right) (\partial^2 u)(\partial v^*)
    \right.
\nonumber\\
    &&\left.
    +\left(12+\frac{3D}{32}\right) (\partial u)(\partial^2 v^*)
    -\left(5+\frac{D}{16}\right) u(\partial^3 v^*)\right],
\nonumber
\eea
and of the term $(\Delta S)_R^{(v)}$ \bref{contribution 4} to
\bea
    \widehat{(\Delta S)}_R^{(v)}=
    \frac{-ik}{6\pi} \int\dif^2 x\,\phi^*_i &&\hspace{-5mm}\left[
     6\partial\left(u (\partial u) (\partial^2\phi^i)\right)
     +\left(9+\frac{3D}{32}\right)
     (\partial^2 u) (\partial u) (\partial\phi^i) \right.
\nonumber\\
     &&\left.
     +\left(8+\frac{2D}{32}\right)
     u (\partial^3 u) (\partial\phi^i) \right].
\nonumber
\eea
In this way, the new one-loop anomaly $\gh A_1$ consists now in the
universal term $(\Delta S)_{R,{\rm univ}}$ \bref{univ anom} plus some
matter dependent contributions.

Radiative corrections induced by the one-loop counterterm $M_1$
\bref{m1 counterterm}, on the other hand,
modify as well the two-loop anomaly, whose explicit
expression is going to be calculated in the next section.
In the nonlocally regularized FA formalism this modification is
conjectured to be described by the two-loop quantum master equation
\be
     \gh A_2=(\Delta M_1)_R +\frac{i}2(M_1,M_1) +i(M_2, S),
\label{twoloop qme}
\ee
with $(\Delta M_1)_R$ given by \bref{reg values} as
$(\Delta M_1)_R\equiv \lim_{\Lambda^2\rightarrow\infty} [\Omega_1]_0,$
and where the explicit expression of $\Omega_1$ is read off from
\bref{delta mp} to be
\be
   \Omega_1=
   \left[(M_1)^A_{\,B}(\delta_\Lambda)^B_{\,C}(\gh\veps^2)^C_{\,A}\right]
   + \left[ S^A_{\,B}
   \left(\delta_\Lambda(\gh O M_{1}) \delta_\Lambda\right)^B_{\,C}
  (\gh\veps^2)^C_{\,A}\right].
\label{omega 1}
\ee
The forthcoming calculation will serve as an explicit verification of this
conjecture at two-loop order.

In the present case, the term $(M_1,M_1)$ in \bref{twoloop qme}
evidently vanishes. Without adding further two-loop counterterms $M_2$, the
two-loop anomaly shift is thus completely contained in
$\Omega_1\sim(\Delta M_1)_R$, for whose explicit computation the quantities
$(M_1)^A_{\,B}$ and $(M_1)_{AB}$, together with
$S^A_{\,B}$, $\gh O^{AB}$, $\gh I_{AB}$ and $(\gh\veps^2)^A_{\,B}$
previously obtained, are needed. $M_1$ \bref{m1 counterterm} being linear
in antifields, it is then clear that the non-vanishing
entries of $(M_1)^A_{\,B}$,
$\{(M_1)^{v^*}_{\,b}, (M_1)^{v^*}_{\,u}, (M_1)^{c^*}_{\,u}\}$,
and $(M_1)_{AB}$,
$\{(M_1)_{ub}, (M_1)_{bu},(M_1)_{uu}\}$,
of which only the explicit expressions of
$\{(M_1)^{v^*}_{\,b}, (M_1)^{c^*}_{\,u}, (M_1)_{ub}\}$
turn out to be necessary in the end for the present calculation
\bea
    (M_1)^{v^*}_{\,b}&=&\beta \left[
    2 u \partial^3 +9 (\partial u)\partial^2
    + 15 (\partial^2 u)\partial + 10 (\partial^3 u)\right]\equiv
    \beta \gh L(u),
\nonumber\\
    (M_1)^{c^*}_{\,u}&=&\beta\left[
    2 u \partial^3 -3 (\partial u)\partial^2
    + 3 (\partial^2 u)\partial -2 (\partial^3 u)\right]\equiv
    -\beta \gh L^\dagger(u),
\nonumber\\
    (M_1)_{ub}&=& -\beta \gh L(v^*),\quad\quad
    (M_1)_{bu}=-(M_1)^\dagger_{ub}= \beta\gh L^\dagger(v^*),
\nonumber
\eea
are antifield independent and linear in antifields, respectively.
Further use of dimensional analysis in terms of the combination $d-j$
\bref{d minus j} and the fact that the integrand of $(\Delta M_1)_R$ should
also have dimension $d=2$ and spin $j=0$, or $d-j=2$, together with
the antifield expansions of the above objects, singles out the
relevant terms in \bref{omega 1} to be
\be
   \Omega_1= \left[\gh\veps^2(M_1)^A_{\,A}\right]+
   \left[\gh\veps^2 (M_1)^A_{\,B}\gh O^{BC}\gh I_{CA} \right]+
   \left[\gh\veps^2 S^A_{\,B}\gh O^{BC}(M_1)_{CA} \right]+
  O\left(\frac{(\Phi^*)^2}{\Lambda^2}\right).
\label{exp omega 1}
\ee

The antifield independent term in \bref{exp omega 1},
describing potential divergencies, vanishes due to the absence of diagonal
terms in $(M_1)^A_{\,B}$. Once again no divergencies appear and
finite contributions to $[\Omega_1]_0$ are seen to come only from the
remaining pieces in \bref{exp omega 1}. The diagonal elements of the
matrices involved in the computation of these terms
\bea
  {\rm diag}\left((M_1)^A_{\,B} \gh O^{BC}\gh I_{CD}\right)&=&
  (0, \,0 , \,B^v_v , \,B^c_c , \,0),
\nonumber\\
  {\rm diag}\left(S^A_{\,B} \gh O^{BC}(M_1)_{CD}\right)&=&
  (0, \,B^b_b , \,0 , \,0 , \,B^u_u),
\nonumber
\eea
results then to be
\bea
  B^b_b&=& 2\beta (u \partial)_{3/2}\, \gh O\,\partial\, \gh L(v^*),
\label{bb entry}\\
  B^v_v &=& 2\beta \gh L(u)\,\gh O\,\partial\, (v^*\partial)_{3/2},
\label{vv entry}\\
  B^c_c &=& -2\beta\gh L^\dagger(u) \,\gh O\,\partial\,
(v^*\partial)_{-1/2},
\label{cc entry}\\
  B^u_u &=& -2\beta (u \partial)_{-1/2}\, \gh O\,\partial \,
  \gh L^\dagger(v^*),
\label{uu entry}
\eea
yielding the following form of $(\Delta M_1)_R$
$$
   (\Delta M_1)_R=
   \lim_{\Lambda^2\rightarrow\infty} {\rm Tr}
  \left[\gh\veps^2\left( B^b_b+ B^v_v +B^c_c +B^u_u\right)\right].
$$

Using then cyclicity of the trace, property
\bref{f partial} and the symmetries of the above expression under the
interchange $u\leftrightarrow v^*$, the above expression can be written in
the more convenient form
$$
   (\Delta M_1)_R=
   \lim_{\Lambda^2\rightarrow\infty} \left\{
   -4\beta\, {\rm Tr} \left[\gh\veps^2\,
  \gh L^\dagger(u)\, \gh O\,\partial\, (v^* \partial)_{-1/2}\right]
   +4\beta\, {\rm Tr} \left[\gh\veps^2\,
  \gh L(u)\, \gh O\,\partial\, (v^* \partial)_{3/2}\right]\right\},
$$
in which the first term groups together the contributions of the spin
2 ghost sector, i.e., the ones coming from the $B^b_b$ \bref{bb entry}
and $B^c_c$ \bref{cc entry} entries, while the second term stands for the
contribution of the spin 3 ghost sector, produced by the $B^v_v$
\bref{vv entry} and $B^u_u$ \bref{uu entry} entries.

Once again, the above functional traces can be computed by means of the
general expression \bref{general trace bis}. In this way, the spin
2 and 3 ghost sector contributions are seen to be, respectively,
$$
   \gh A_2^{(2)}=\frac{i\beta}{5\pi} \frac{199}{8}
   \int\dif^2 x\,u\,\partial^5 v^*,\quad\quad
   \gh A_2^{(3)}=\frac{i\beta}{5\pi} \frac{149}{8}
   \int\dif^2 x\,u\,\partial^5 v^*,
$$
which add up, after substituting the actual value of $\beta$,
to the one-loop correction $\gh A_2$ \bref{twoloop qme} produced by the
counterterm $M_1$ \bref{m1 counterterm} to the two-loop anomaly
\be
    \gh A_2= \frac{-i 87 Dk}{1920\pi^2}
   \int\dif^2 x\,u\,\partial^5 v^*,
\label{2loop shift}
\ee
in complete agreement, apart from numerical factors due to the difference
in the used conventions, with previous results \ct{hull93}. This
computation constitutes thus a direct verification for $p=2$ of the
conjecture stated along the paper, namely, of the interpretation of
expression \bref{reg higher order} for the higher order terms of the
quantum master equation as one-loop corrections to the $p$-loop anomaly
generated by the counterterms $M_k$, $k<p$ and, consequently, of the
incompleteness of the regulated quantum master equation
\bref{final reg qme} itself.

\subsection{Universal two-loop Anomaly}

\hspace{\parindent}
In this section, in order to finally show that incompleteness of the
quantum master equation comes from a naive derivation of the FA BRST Ward
identity \bref{ward identity}, or of its regulated version
\bref{reg ward ident}, and not as a drawback of the nonlocal regularized
FA formalism by itself, the
computation of the universal $W_3$ two-loop anomaly is performed by using
the ``left hand side'' of the BRST Ward identity \bref{reg ward ident}
that is, by first calculating the relevant part of the two-loop effective
action in the nonlocally regulated FA framework and afterwards performing
its BRST variation.

The relevant two-loop 1PI diagram,
involving only matter fields $\phi^i$ in its internal lines and the
antifields $v^*$ ($v^*=-B$, when in the classical basis) as external
sources (Fig.\,2)

\begin{picture}(20000,10000)

\drawline\photon[\E\REG](16500,5000)[3]
\put(\pmidx,6000){$v^*$}
\global\advance\pbackx by 1500
\put(\pbackx,\pbacky){\circle{3000}}
\global\advance\pbackx by -1500
\drawline\fermion[\E\REG](\particlebackx,\particlebacky)[3000]
\drawline\photon[\E\REG](\particlebackx,\particlebacky)[3]
\put(\pmidx,6000){$v^*$}

\end{picture}

\centerline{Fig.\,2. Unregularized two-loop matter diagram.}

\vskip 0.7cm

\noindent
comes from the contraction of the vertex
$-\frac13 v^* d_{ijk} (\partial\phi^i) (\partial\phi^j) (\partial\phi^k)$
of the gauge fixed proper solution \bref{w3gfa} with itself.
Its unregularized contribution is read off from the Feynman rules obtained
from \bref{w3gfa} to be
\be
   \Gamma(v^*, v^*)=-\frac{i}3 \hbar^2 d^2\int\dif^2 x\,\dif^2 y\,\,
   v^*(x)
   \left[\partial\left(\frac{i}{\partial\bar\partial}\right)\partial\,
   \delta^2(x-y)\right]^3
   v^*(y),
\label{unreg gamma 2}
\ee
with $d^2\equiv d_{ijk} d^{ijk}$.

As previously noticed in the original references \cite{kw92,kw93},
when performing diagramatic calculations, the nonlocally regulated theory
is effectively realized by using the auxiliary action \bref{auxiliary proper}
and eliminating by hand closed loops formed solely with barred lines.
For the present calculation, the relevant terms in the auxiliary proper
solution associated with \bref{w3gfa}, giving the modified propagators and
the ``regularized'' version of the considered vertex, are
\bea
   \tilde S&=&\int\dif^2 x\,\left[
   \frac12 \phi^i\left(\frac{\partial\bar\partial}{\gh\veps^2}\right)\phi^i
   +\frac12
   \psi^i \left(\frac{\partial\bar\partial}{1-\gh\veps^2}\right)\psi^i
   \right.
\nonumber\\
   &&\left.
   -\frac13 (\gh\veps^2 v^*) d_{ijk}
  (\partial(\phi^i+\psi^i)) (\partial(\phi^j+\psi^j))
  (\partial(\phi^k+\psi^k))+\ldots\right],
\nonumber
\eea
where $\psi^i$ stand for the shadow fields corresponding to the original
matter fields $\phi^i$. The new Feynman rules coming from this action lead
then to the following $2^3=8$ diagrams with only internal matter lines
and two external $v^*$ lines (Fig.\,3)

\begin{picture}(50000,10000)


\drawline\photon[\E\REG](-1500,5000)[3]
\global\advance\pbackx by 1500
\put(\pbackx,\pbacky){\circle{3000}}
\global\advance\pbackx by -1500
\drawline\fermion[\E\REG](\particlebackx,\particlebacky)[3000]
\drawline\photon[\E\REG](\particlebackx,\particlebacky)[3]


\drawline\photon[\E\REG](11000,5000)[3]
\global\advance\pfrontx by -1600
\put(\pfrontx,5000){$3\times$}
\global\advance\pfrontx by 1600
\global\advance\pbackx by 1400
\put(\pbackx,6000){$\rule{1mm}{3mm}$}
\global\advance\pbackx by -1400
\global\advance\pbackx by 1500
\put(\pbackx,\pbacky){\circle{3000}}
\global\advance\pbackx by -1500
\drawline\fermion[\E\REG](\particlebackx,\particlebacky)[3000]
\drawline\photon[\E\REG](\particlebackx,\particlebacky)[3]


\drawline\photon[\E\REG](23500,5000)[3]
\global\advance\pfrontx by -1600
\put(\pfrontx,5000){$3\times$}
\global\advance\pfrontx by 1600
\global\advance\pbackx by 1400
\put(\pbackx,6000){$\rule{1mm}{3mm}$}
\global\advance\pbackx by -1400
\global\advance\pbackx by 1500
\put(\pbackx,\pbacky){\circle{3000}}
\global\advance\pbackx by -1500
\drawline\fermion[\E\REG](\particlebackx,\particlebacky)[3000]
\global\advance\pmidx by -100
\put(\pmidx,4500){$\rule{1mm}{3mm}$}
\global\advance\pmidx by 100
\drawline\photon[\E\REG](\particlebackx,\particlebacky)[3]


\drawline\photon[\E\REG](34500,5000)[3]
\global\advance\pbackx by 1400
\put(\pbackx,6000){$\rule{1mm}{3mm}$}
\put(\pbackx,3200){$\rule{1mm}{3mm}$}
\global\advance\pbackx by -1400
\global\advance\pbackx by 1500
\put(\pbackx,\pbacky){\circle{3000}}
\global\advance\pbackx by -1500
\drawline\fermion[\E\REG](\particlebackx,\particlebacky)[3000]
\global\advance\pmidx by -100
\put(\pmidx,4500){$\rule{1mm}{3mm}$}
\global\advance\pmidx by 100
\drawline\photon[\E\REG](\particlebackx,\particlebacky)[3]

\end{picture}

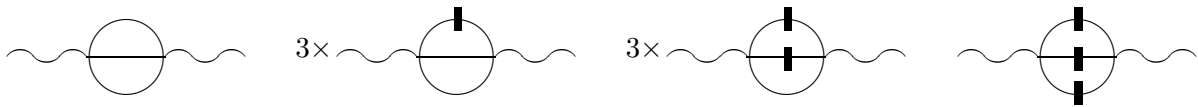
\centerline{Fig.\,3. Two-loop matter diagrams in
                     the nonlocally regulated theory.}

\vskip 0.7cm
\noindent
where, in the second and third diagrams ``$3\times$'' stands for the three
different diagrams obtained by combining the corresponding unbarred and
barred propagators.

{}From them, and upon elimination of the last contribution formed solely
with barred lines, the regularized expression of $\Gamma(v^*, v^*)$
\bref{unreg gamma 2} arises
\be
   \Gamma_{\Lambda}(v^*, v^*)=
   -\frac{i}3 \hbar^2 d^2\int\dif^2 x\,\dif^2 y\,
   \left(\gh\veps^2 v^*(x)\right)
   K(x,y;\Lambda^2)
   \left(\gh\veps^2 v^*(y)\right),
\label{reg gamma 2}
\ee
where the regularized expression of the kernel $K(x,y;\Lambda^2)$ in
\bref{reg gamma 2} reads, using the more convenient momentum
representation
\bea
   &&K(x,y;\Lambda^2)=
   \int\frac{\dif^2 p}{(2\pi)^2}\, e^{ip\cdot(x-y)}
   \left\{(-i)^3
   \int_{R_1^3}\frac{\dif t_1\dif t_2 \dif t_3}{\Lambda^6}
   \int\frac{\dif^2 k_1}{(2\pi)^2} \int\frac{\dif^2 k_2}{(2\pi)^2}
   \right.
\nonumber\\
   &&\left.
   (ik_1)^2\,(ik_2)^2\, (i(p-k_1-k_2))^2
   \exp\left[ -t_1\frac{k_1\bar k_1}{\Lambda^2}
   -t_2\frac{k_2\bar k_2}{\Lambda^2}
   -t_3\frac{(p-k_1-k_2)(\bar p-\bar k_1-\bar k_2)}{\Lambda^2}\right]
    \right\},
\nonumber
\eea
with $R_1^3$ denoting the standard 3-parameter region of integration at
two-loops, namely, the first quadrant of $R^3$ minus the cube $[0,1]^3$
\cite{kw93}.

The term in the exponential can now be ``diagonalized'', and
integrations over momenta decoupled,
by performing the standard change of variables
\bea
   &&k^\mu_1\,\rightarrow\,
   \left[\frac{\Lambda}{(t_1+t_3)^{1/2}}\, k^\mu_1
         -\frac{\Lambda\, t_3}{(t_1+t_3)^{1/2}\,t_{123}^{1/2}}\, k^\mu_2
         + \frac{t_2 t_3}{t_{123}}\, p^\mu\right],
\nonumber\\
   &&k^\mu_2\,\rightarrow\,
   \left[\frac{\Lambda(t_1+t_3)^{1/2}}{t_{123}^{1/2}}\, k^\mu_2
         + \frac{t_3 t_1}{t_{123}}\, p^\mu\right],
\nonumber
\eea
with the combination $t_{123}$ defined as
$t_{123}=(t_1t_2+t_2t_3+t_3t_1)$.
Afterwards, in the polynomial in $k_1$, $k_2$ arising upon this change
from the factor
$\left[(ik_1)^2 (ik_2)^2 (i(p-k_1-k_2))^2\right]$ in the
integrand, only the $k$-independent term survives, as deduced from
integrals \bref{momentum int} in the appendix. Performing then the
remaining gaussian integrals in $k_1$, $k_2$, the following result arises
\be
   K(x,y;\Lambda^2)=
   (-i)^3 \left(\frac{-i}{2\pi}\right)^2
   \int\frac{\dif^2 p}{(2\pi)^2} \frac{(ip)^6}{\Lambda^2}\,
   e^{ip\cdot(x-y)} \gh K\left(\frac{p\bar p}{\Lambda^2}\right),
\label{kernel}
\ee
with the quantity $\gh K\left(\frac{p\bar p}{\Lambda^2}\right)$ given by
\be
   \gh K\left(\frac{p\bar p}{\Lambda^2}\right)=
   \int_{R_1^3} \dif t_1\dif t_2 \dif t_3\,
   \frac{(t_1 t_2 t_3)^4}{t_{123}^7}
   \exp\left[- \frac{t_1 t_2 t_3}{t_{123}}\,
   \frac{p\bar p}{\Lambda^2}\right].
\label{int parameters}
\ee

This type of integrals are most easily performed by using the general
change of variables in parameter space depicted in \cite{kw93}. In the
present case, that change amounts to
$(t_1,\, t_2,\, t_3)\,\rightarrow\, (x,\,y,\,t)$
with $x$, $y$ and $t$ expressed as
$$
  t= t_1+t_2+t_3,\quad\quad
  x= \frac{t_2+t_3}{t},\quad\quad
  y= \frac{t_3}{t}.
$$
In terms of these variables, the integral \bref{int parameters} becomes
\bea
   \gh K\left(\frac{p\bar p}{\Lambda^2}\right)&=&
   6 \int_0^{1/3}\dif y \int_{2y}^{(1+y)/2}\dif x
   \frac{\left[(1-x)(x-y) y\right]^4}
   {\left[x(1-x)+y(x-y)\right]^7}
\nonumber\\
  && \times \int_{1/(x-y)}^{\infty}\dif t\,
  \exp\left[-t\left(\frac{(1-x)(x-y) y}{x(1-x)+y(x-y)}\,
   \frac{p\bar p}{\Lambda^2}\right)\right],
\nonumber
\eea
so that the integral over $t$ is easily performed, producing
an incomplete gamma function, defined as
$$
   \Gamma(n,z)= \int_{z}^{\infty}\dif t\, t^{n-1}\, e^{-t},
$$
and yielding
\bea
   \gh K\left(\frac{p\bar p}{\Lambda^2}\right)&=&
   6 \int_0^{1/3}\dif y \int_{2y}^{(1+y)/2}\dif x\,
   \frac{\left[(1-x)(x-y) y\right]^3}
   {\left[x(1-x)+y(x-y)\right]^6}\,
   \frac{\Lambda^2}{p\bar p}
\nonumber\\
   &&\times\Gamma\left(1, \frac{y (1-x)}{x(1-x)+y(x-y)}\,
   \frac{p\bar p}{\Lambda^2}\right).
\label{int parameters 2}
\eea

Upon substitution now of \bref{int parameters 2}
in the expression of the regularized kernel
\bref{kernel}, the respective $\Lambda^2$ and $\Lambda^{-2}$ factors
cancel out. The resulting expression is then finite, so that the limit
$\Lambda^2\rightarrow\infty$, which gives $\Gamma(1,0)=\Gamma(1)=1$ for the
incomplete gamma function in \bref{int parameters 2}, can be safely taken,
yielding
$$
   \lim_{\Lambda^2\rightarrow\infty} K(x,y;\Lambda^2)=
   \frac{3 a i}{2\pi^2} \int\frac{\dif^2 p}{(2\pi)^2}\,
   \frac{(ip)^6}{(ip)(i\bar p)}\, e^{ip\cdot(x-y)} =
   \frac{3 a i}{2\pi^2}
   \frac{\partial^5}{\bar\partial}\,\delta^2(x-y),
$$
with the numerical factor $a$ given by the integral
$$
   a= \int_0^{1/3}\dif y \int_{2y}^{(1+y)/2}\dif x
   \frac{\left[(1-x)(x-y) y\right]^3}
   {\left[x(1-x)+y(x-y)\right]^6}.
$$
The further change of variables $\{u=1/x,\, v=y/x\}$, followed by the
subsequent change $u\rightarrow(u-1)/v(1-v)$, brings the above integral to
a simpler form, giving finally the result
$$
   a= \int_0^{1/2}\dif v \,v(1-v) \int_{1/v}^{\infty}\dif u
   \frac{u^3}{(1+u)^6}= \frac1{720}.
$$

All in all, taking the above results into account, the following
regularized expression of $\Gamma(v^*, v^*)$ \bref{reg gamma 2},
in the limit $\Lambda^2\rightarrow\infty$, comes out
\be
   \Gamma(v^*, v^*)=\frac{\hbar^2 d^2}{1440\pi^2}
   \int\dif^2 x\, v^*\, \frac{\partial^5}{\bar\partial}\, v^*,
\label{fin reg gamma}
\ee
whereas further calculation of its BRST variation, by using the
transformation
$$
  \delta v^* = (v^*, S)=-\bar\partial u + O(\Phi^*),
$$
finally yields, after nonlocal radiative corrections induced by the
one-loop anomaly are separated off, the universal two-loop anomaly
\be
   \gh A_2(v^*, u)=\frac{i d^2}{720\pi^2}
   \int\dif^2 x \, u \,\partial^5 \,v^*,
\label{two loop anom}
\ee
to which expression \bref{2loop shift} should be added when considering the
one-loop counterterm $M_1$ \bref{m1 counterterm} included in the theory.

Expressions \bref{fin reg gamma} for (part of) the two-loop effective
action and \bref{two loop anom} for the universal two-loop anomaly,
are in agreement
--apart from numerical factors which can be traced back to the
diferent form of the actions used in the computations--
with earlier results in the literature
\cite{mat89,hull91,ssn91,prs91,hull93}. This fact indicates thus that,
at least for this case, the nonlocally regularized FA formalism is able to
deal with two-loop corrections and reproduce the correct numerical
coefficients for the anomalies.

\section{Conclusions and Perspectives}

\hspace{\parindent}%
The aim of this paper has been to extend to general gauge theories
the nonlocal regularization method of ref.\,\ct{emkw91,kw92,kw93},
by reformulating it according to the ideas of the antibracket-antifield
formalism. As a result of this procedure, a nonlocal regularized form
$S_\Lambda$ \bref{nonlocal proper}
of the proper solution of the classical master equation $S$ \bref{gfps}
comes out, which encodes all the information about the structure of the
regulated BRST symmetry. Further extension of the basic ideas of nonlocal
regularization at quantum level provides a systematic way to construct the
nonlocal regularized version $W_\Lambda$ \bref{w lambda} of the quantum
action $W$ \bref{quantum action}. In the end, its use in the basic path
integral leads to a fully
regularized version of the FA formalism, which is able to treat higher
order loop corrections and offers a suitable framework for the analysis
of the BRST Ward identity \bref{reg ward ident} and of the BRST anomaly
\bref{fin anomaly 2} proposed by FA through the quantum master equation.

{}From the analysis of this fundamental equation, based in the study of
the action of the operator $\Delta$ \bref{delta op} on the regularized
quantum action $W_\Lambda$ \bref{w lambda}, and supplemented with explicit
calculations performed in the
example of chiral $W_3$ gravity, one of the main conclusions of this
paper arises, namely, the incompleteness of the quantum
master equation and its ability to describe only the one-loop
corrections generated by the counterterms $M_k$, $k<p$, to the $p$-loop
anomalies. In other words, the present form of the quantum master equation
seems to be the ``one-loop form'' of a
general expression, presumably defined in terms of suitable quantum
generalizations of the operator $\Delta$ \bref{delta op},
$\Delta_q=\Delta+\hbar\Delta_1+\ldots$, and of
the antibracket \bref{antibracket} involved in its definition.

The incompleteness of the present form of the quantum master equation,
on the other hand, can
be traced back to the naive character of the FA derivations in which this
quantity appears as obstruction. Indeed, apart from threatening the
fulfillment of the BRST Ward identity, the quantum master equation also
controls, for instance, the theory's (in)dependence of the gauge choice
\ct{bv81}, as the following equation shows
\be
  Z_{\Psi+\delta\Psi}= Z_{\Psi}+
  \int\gh D \Phi\exp\left[\frac{i}{\hbar} W_\Sigma(\Phi)\right]
   \left[\Delta W+\frac{i}{2\hbar}(W,W)\right]_\Sigma\cdot\delta\Psi,
\label{gauge dep}
\ee
where the subindex $\Sigma$ stands for the restricition to the
so-called gauge fixing surface, defined in terms of a suitable
gauge-fixing fermion $\Psi$ as $\Sigma=\{\Phi^*=\der{\Psi}{\Phi}\}$, and
with $Z_{\Psi}$ the value of the generating functional
\bref{generating funct} in this surface when $J=0$. This equation is
derived by first varying infinitesimaly the gauge fermion $\Psi$ in
$Z_{\Psi}$ to $\Psi+\delta\Psi$, and evaluating afterwards the resulting
expression by considering the invariance of $Z_\Psi$ under the
change of variables $\Phi\rightarrow\Phi+\delta\Phi$, with $\delta\Phi$
given by
\be
  \delta\Phi^A=(\Phi^A,W)_\Sigma\cdot\delta\Psi.
\label{brst change}
\ee
In this way, the term $\Delta W$ in \bref{gauge dep} originates in the
noninvariance of the functional measure under \bref{brst change}, while
$(W,W)$ comes from the transformation of the integrand.

However, the treatment of field redefinitions at functional level in this
way, namely, by merely changing integration variables in functional
integrals, has been known to be incorrect for a long time. Indeed, in
ref.\,\ct{gj76}, nonlinear point canonical transformations were analyzed
at the quantum mechanical level, using a discretized version of the path
integral, and the original and naively transformed theory were found to
differ by terms of order $\hbar^2$. More recently, this question has been
analyzed as well at the field theoretical level in \ct{kc94} on the
same grounds, with the result that, upon making nonlinear field
redefinitions, extra terms of
order $\hbar^2$ and higher generally appear together with the ones expected
to come from the naive change of variables.

In view of these results and taking into account the nonlinear character
of the BRST transformation
\bref{brst change} in general, it is natural to expect that the missing
$O(\hbar^2)$ terms in the present form of the quantum master equation,
the ones which shall characterize higher order loop anomalies,
should come precisely from the extra $O(\hbar^2)$ pieces generated
upon performing the change of variables \bref{brst change}, i.e., upon
making in the proper way changes of variables in the path integral.
Implementation of these ideas in the nonlocally regulated FA formalism,
in order to translate to the continuum the lattice expressions obtained in
\ct{kc94}, would then be useful in order to explicitly determine the
form of the higher order terms involved in the
complete form of the quantum master equation and, as a consequence, of the
higher order loop anomalies, without relying on diagramatic calculations.
In the end, such expressions would also possibly allow an algebraic
characterization of higher order loop anomalies, in much the same
way as one-loop anomalies are characterized as being non-trivial
BRST cocycles at ghost number one. Work towards this algebraic approach to
characterize anomalies by means of a ``complete'' quantum master equation
is already in progress.

\section*{Acknowledgements}

\hspace{\parindent}%
It is a pleasure to thank J.\,Gomis, W.\,Troost, A.\,van Proeyen,
F.\,Brandt, F.\,de Jonghe, R.\,Siebelink and S.\,Vandoren for many fruitful
discussions.
Financial support from the spanish ministry of education (MEC) is also
acknowledged.

\appendix

\section{Appendix: Functional Traces}

\hspace{\parindent}%
This appendix is devoted to the computation of the general form of the
functional traces involved in the computations performed in section 4.
The first of these traces appears in the evaluation of the divergent
term of $\Omega_0$ \bref{antifield exp delta}
$$
   \gh T(F,n)=
   {\rm Tr}\left[\gh\veps^2(F(\Phi,\Phi^*)\,\partial)_{n}\right],
$$
with the operators $\gh\veps^2$ and $(F(\Phi,\Phi^*)\,\partial)_{n}$
defined by \bref{w smearing op} and \bref{f partial}, respectively.
Introducing as usual momentum space eigenfunctions to saturate the sum,
the above trace can be written as
\bea
   \gh T(F,n)&=&
   \int\dif^2 x\,\int\frac{\dif^2 k}{(2\pi)^2}\,
   e^{-ik\cdot x}\left\{[F\partial+n(\partial F)]
   \exp\left(\frac{\partial\bar\partial}{\Lambda^2}\right)\right\}
   e^{ik\cdot x}
\nonumber\\
   &=& \int\dif^2 x\,\int\frac{\dif^2 k}{(2\pi)^2}\,
   \left[ik\, F +n(\partial F)\right]
   \exp\left(\frac{-k\bar k}{\Lambda^2}\right).
\nonumber
\eea
Further rescaling of the momentum $k$ by $\Lambda$ and Wick rotation to
euclidian space, in order to obtain the usual gaussian damping factors,
finally yields
\be
   \gh T(F,n)= -\Lambda^2\,\frac{i\,n}{2\pi}
   \int\dif^2 x\, (\partial F)=0,
\label{integral 0}
\ee
as a consequence of the integrals
\be
   \int\frac{\dif^2 k}{(2\pi)^2}\, \exp\left(-k\bar k\right)=
   \frac{-i}{2\pi},\quad\quad
   \int\frac{\dif^2 k}{(2\pi)^2}\, (k\stackrel{n}{\ldots}k)\,
   \exp\left(-k\bar k\right)=0,
   \quad\quad\forall\, n>0,
\label{momentum int}
\ee
and the fact that the integrand in \bref{integral 0} is a total derivative.

The second relevant integral, useful for the calculation of finite
contributions to $\Omega_0$ \bref{antifield exp delta} and $\Omega_1$
\bref{exp omega 1}, is given by the general expression \bref{general trace}
$$
  \gh T(F,G; n,m)= \lim_{\Lambda^2\rightarrow\infty}
  \gh T(F,G; n,m; \Lambda^2)=
  \lim_{\Lambda^2\rightarrow\infty}
  {\rm Tr} \left[\gh\veps^2\,
  F\,\partial^n \,\gh O\partial\, G\,\partial^m\right],
  \quad\mbox{for} \quad n,m\geq 0.
$$
Upon use of momentum space eigenfunctions,
as in the previous computation, and further introduction of
the momentum representation
$G(x)= \int\frac{\dif^2 p}{(2\pi)^2} G(p) e^{ip\cdot x}$ yields
\bea
  &&\gh T(F,G; n,m;\Lambda^2)=
  \int\dif^2 x\, F(x) \left\{
   \int\frac{\dif^2 k}{(2\pi)^2} \int\frac{\dif^2 p}{(2\pi)^2}\right.
\nonumber\\
   &&\left.\left[ \int_0^1\frac{\dif t}{\Lambda^2}
   \exp\left(-t\frac{(k+p)(\bar k+\bar p)}{\Lambda^2}
   -\frac{k\bar k}{\Lambda^2}\right)\right]
   (ik)^m\, (ip+ik)^{n+1} G(p) e^{ip\cdot x}\right\},
\label{11 before limit}
\eea
The integrals over $k$ and $p$ can now be decoupled by performing the
standard change of variables
$$
   k^\mu\,\rightarrow\,
   \left(\frac{\Lambda}{\sqrt{1+t}}\,k^\mu-\frac{t}{1+t}\,p^\mu \right).
$$
The factor $[(ik)^m(ip+ik)^{n+1}]$ in the integrand of
\bref{11 before limit} produces then a polynomial in $k$,
of which only the $k$-independent term survives, as a consequence of
\bref{momentum int}. The result of these manipulations is thus
\bea
  &&\gh T(F,G; n,m;\Lambda^2)=
\nonumber\\
  &&\frac{-i}{2\pi}
   \int\dif^2 x\, F(x)\left\{
   \int\frac{\dif^2 p}{(2\pi)^2}\left[
   \int_0^1\dif t\,\frac{(-1)^m\, t^m}{(1+t)^{n+m+2}}
   \exp\left(-\frac{t}{1+t}\,\frac{p\bar p}{\Lambda^2}\right)\right]
   (ip)^{n+m+1} G(p) e^{ip\cdot x}\right\}=
\nonumber\\
   &&\frac{-i}{2\pi}
   \int\dif^2 x\, F\, \bar\gh O(n,m;\Lambda^2)\,\partial^{n+m+1} G,
\nonumber
\eea
where the factor $-i/2\pi$ comes from integration over the internal
momentum $k$, and where the operator $\bar\gh O(n,m;\Lambda^2)$ is defined
by
$$
   \bar\gh O(n,m;\Lambda^2)=
   \int_0^1\dif t\,\frac{(-1)^m\,t^m}{(1+t)^{n+m+2}}
   \exp\left(\frac{t}{1+t}\,\frac{\partial\bar\partial}{\Lambda^2}\right).
$$
The limit $\Lambda^2\rightarrow\infty$ of the above operator is well
defined, yielding
\bea
  \bar\gh O(n,m)&=& \lim_{\Lambda^2\rightarrow\infty}
   \bar\gh O(n,m;\Lambda^2)=
   \int_0^1\dif t\,\frac{(-1)^m\,t^m}{(1+t)^{n+m+2}}
\nonumber\\
   &=& \sum_{k=0}^m {m \choose k}
   \frac{(-1)^k}{n+m+1-k}\left(1-\frac1{2^{n+m+1-k}}\right),
\label{onm}
\eea
and leading to the final result \bref{general trace bis}
$$
  \gh T(F,G; n,m)= \frac{-i}{2\pi}\,\bar\gh O(n,m)\,
  \int\dif^2 x\, F\,\partial^{n+m+1}\, G,
  \quad \mbox{for} \quad n,m\geq 0,
$$
with $\bar\gh O(n,m)$ given by \bref{onm}.

\endsecteqno

\end{document}